\documentclass[onecolumn]{aa}
\usepackage{natbib}
\usepackage{graphicx}
\usepackage{graphics}
\usepackage{longtable}
\usepackage{lscape}
\usepackage{amssymb}
\usepackage{txfonts}
\bibstyle{aa} 
\def\arcsec{\ifmmode^{\prime\prime}\;\else$^{\prime\prime}\;$\fi}

\begin{document}
   \title{High redshift X--ray galaxy clusters. II. The L$_{X}-$T 
          relationship revisited}
   \author{M. Branchesi
           \inst{1,2}, 
           I. M. Gioia
           \inst{2}, 
           C. Fanti
           \inst{2}, 
           R. Fanti
           \inst{2}}

   \offprints{M. Branchesi}

   \institute{Dipartimento di Astronomia, Universit\`a di Bologna,
              Via Ranzani 1, 40127 Bologna, Italy \\
                \email{m.branchesi@ira.inaf.it}
                           \and
            Istituto di Radioastronomia, INAF, Via Gobetti 101, 
            40129 Bologna, Italy \\
                \email{gioia@ira.inaf.it,
                      rfanti@ira.inaf.it, cfanti@ira.inaf.it}}  

   \date{Received 13 March 2007; accepted 17 June 2007}

\abstract{}
{In this paper we re-visit the observational relation between  X--ray
luminosity and temperature for high-z galaxy clusters and
compare it with the local $\rm L_{X}$--T and with theoretical models.}
{To these ends we use a sample of 17 clusters extracted from the 
\textit{Chandra} archive supplemented with additional clusters from the
literature, either observed by  \textit{Chandra} or XMM--\textit{Newton}, 
to form a final sample of 39 high redshift (0.25 $<$ z $<$1.3) objects.
Different statistical approaches are adopted to analyze the $\rm L_{X}$--T
relation.}
{The slope of the $\rm L_{X}$--T relation of high redshift clusters is
steeper than expected from the self--similar model predictions and
steeper, even though still compatible within the errors, than the
local $\rm L_{X}$--T slope. The distant cluster $\rm L_{X}$--T
relation shows a significant evolution with respect to the local
Universe: high-z clusters are more luminous than the local ones by a
factor $\approx 2$ at any given temperature. The evolution with
redshift of the L$_{X}$--T relation cannot be described by a single
power law nor by the evolution predicted by the self--similar model.}
{We find a strong evolution, similar or stronger than the
self--similar model, from z = 0 to z $\le 0.3$ followed by a much
weaker, if any, evolution at higher redshifts. The weaker 
evolution is compatible with non-gravitational models of structure
formation. According to us a statistically significant sample of
nearby clusters (z$<$0.25) should be observed with the current
available X-ray telescopes to completely exclude observational effects
due to different generation detectors and to understand this novel
result.}


\authorrunning{Branchesi et al.}
\titlerunning{The L$_{X}-$T relationship revisited}

\maketitle

\section{Introduction}

Clusters of galaxies represent the largest scale of fully collapsed
structures in the Universe and thus offer a unique insight into the
formation of structures and into the parameters governing their
evolution.  As powerful X-ray emitters ($\rm L_{X} = 10^{43}-10^{45}
erg ~s^{-1}$) galaxy clusters can be easily traced up to high
redshift, hence they are a very important observational tool for
cosmologists.  One problem that must be examined is how to relate the
observable quantities (e.g., cluster X-ray gas temperature or
luminosity) to the quantity predicted by the cosmological studies
(usually mass).  A detailed and accurate knowledge of the relations
between observable bulk features is necessary. For instance, the
relation between the X--ray luminosity and temperature of the
intra-cluster medium (ICM), that is analyzed in detail in this paper,
provides a means to convert an easily observed X-ray luminosity
function into a more cosmologically useful mass function
\citep[e.g.,][]{Ev96,Rei02,Ev02,Al03,Bo06}.

\smallskip\noindent 
Under the assumption that the baryons follow the total matter
distribution (dominated by dark matter) \cite{Ka86} constructed the
simple self--similar model of cluster formation where the ICM is
driven only by gravitational processes like shocks and adiabatic
compression.  As long as the baryons are distributed in the same way
as the total mass, each X--ray observable scales like some power of
the mass.  The self--similar model, supported by N--body
hydrodynamical simulations which consider only gravitational processes
\citep{Bry98}, predicts the relation between luminosity and
temperature to be $\rm L_{X} \propto T^2$.

\smallskip\noindent 
The observational tight correlation between luminosity and temperature
found for low redshift cluster samples \citep[e.g][]{Ma98,Ar99}
indicates a similar formation history for all clusters and is
qualitatively similar to the predictions of the self--similar models
of the cluster formation. However the observed slope of the L$_{X}$--T
relation is found to be much steeper ($\simeq 3$) than expected from
the self--similar model.  Such deviation from self--similarity is
taken as evidence that the simple gravitational collapse is not the
only process that governs the heating of baryons, and that
non-gravitational processes occurring before or during the cluster
formation contribute to the clusters energy budget in a non negligible
way. To explain the departure from self--similarity an increase in the
central entropy is required \citep[e.g.] {Ev91,Ka91,Po99,Tn01}.  This
extra entropy makes the gas harder to compress in the cluster core,
particularly in the shallower potential wells of low temperature
clusters, whose thermal bremsstrahlung emission is reduced as compared
to the self--similar model prediction. The physical processes that can
rise the entropy of the gas involve heating by supernovae and by
active galactic nuclei (AGNs)\citep[e.g.][]{Bal99,Bow01,BM01,Bor02},
or the removal of low-entropy gas via cooling
\citep[e.g.][]{Bry00,Vo01,Wu02} or a combination of heating and
radiative cooling \citep[e.g.][]{Vo02,Tor03,Bo04,Mu06}.  Therefore an
accurate analysis of the $\rm L_{X}-T$ relationship is also important
to investigate the physics behind galaxy cluster formation.

\smallskip\noindent
Further information on the formation and evolution of structure in 
the Universe are given by the study of the statistical properties of high
redshift cluster samples with respect to the low redshift samples. Any
observed change in the L$_{X}$--T scaling law at high redshift with
respect to low redshift allows in principle to verify the predictions
of the cluster self--similar model or to constrain different scenarios
of heating schemes and cooling efficiency \citep[e.g.][]{Vo05a}.

\smallskip\noindent
With the main goal to revisit the L$_{X}$--T relationship for high-z
clusters we used seventeen clusters in the range 0.25 $<$ z $<$ 1.01
selected from the {\em Chandra} archive. We analyzed the data as
described in details in a companion paper \citep[][Paper I]{Br07a}.  
In order to increase the statistics the sample was
supplemented with additional clusters from the literature with
redshift up to z$\sim$1.3, either observed by {\em Chandra} or
XMM-{\em Newton}.  The final sample consists of 39 objects. The
observational results from the L$_{X}$--T analysis of the 39 distant
clusters are compared with those obtained from the analysis of the
L$_{X}$--T relation of low-z clusters \citep{Ma98,Ar99} and 
with theoretical models for structure formation.

\smallskip\noindent
All the uncertainties in the paper are at the 1 $\sigma$ confidence
level, unless otherwise noted.  We use a $\Lambda$CDM cosmology with
$\rm H_{0} = 70 ~km ~s^{-1} ~Mpc^{-1}$ and $\Omega_{\rm m} = 
1-\Omega_{\Lambda} = 0.3$.

\section{Cluster Scaling Relations}
\label{LTtheory}

%----------------------------- subsection ----------------------------
\subsection{The self--similar model}
\label{selfsim}

The self--similar model detailed by \cite{Ka86} is the most natural
expectation in a picture where clusters form via the collapse of the
most dense regions and cluster baryons are heated only by
gravitational processes (adiabatic compression and shock heating)
during the collapse. In this scenario it is assumed that a constant
fraction of the baryonic content of the Universe ends up in the
intracluster gas in hydrostatic equilibrium within the dominant dark
matter, and emits by thermal bremsstrahlung.
A state of {\it virial equilibrium} is achieved within a {\it virial
radius} $R_{\rm v}(z)$ which includes a mean density which is a factor
${\Delta}_{\rm v}(z)$ above the critical density of the Universe, 
$\rho_{c,z}= 3 H_z^2/(8 \pi G)$, at that redshift.  A cosmology factor 
$E_z$ is used hereafter to indicate the evolution of the Hubble constant 
at redshift z with respect to z$=$0 for a flat cosmology with matter 
density $\Omega_{\rm m}$:
\begin{eqnarray} 
E_z = H_z / H_0 = \left[\Omega_{\rm m} (1+z)^3 + 1 - \Omega_{\rm m} 
                      \right]^{1/2}
\label{eqn1}
\end{eqnarray}
In a flat $\Lambda = 0$ cosmology  $E_z = (1+z)^{1.5}$.

\smallskip\noindent
The density contrast is given by
\begin{eqnarray} 
{\Delta}_{\rm v}(z) = 18\pi^2 + 82 (\Omega_{z}-1) - 39  
                          (\Omega_{z}-1)^2  ~~~~~~\rm {(Bryan~\&~Norman~1998)}
\label{eqn2}
\end{eqnarray}
\noindent
with $\Omega_{z} = \Omega_{m} (1+z)^3/ E_z^2$. In a flat $\Lambda = 0$
cosmology $\Delta_{\rm v}$ = 18$\pi^2$ and it is independent of
redshift.
 
\smallskip\noindent
In this simple scenario, called {\it the self--similar scenario}, the
properties of clusters of different masses and at different redshifts
are related to one-another according to simple scaling laws.  The
expected properties of clusters at high redshift are identical to
those of low-redshift clusters, apart from scaling factors
reflecting the increase of the mean density of the Universe with
redshift. The self--similar scenario predicts how the ICM physical
properties, like luminosity and temperature, are related
\citep[e.g.][]{Ka86,Ka91,Ev91}.

\noindent
Under the assumption that the ICM is spherically symmetric, for a
$\beta$-model \citep{Ca76} the total mass within a radius R is given 
by the following equation \citep[see e.g.][]{Vo05a}:
\begin{eqnarray} 
\rm M_{tot} (< R) = \frac {3 \beta k_{B} T(^{\circ} K) R_{c}}{G\mu m_{p}} 
\frac{(R/R_{c})^3}{1+(R/R_{c})^2}
\label{eqn3}
\end{eqnarray} 
where the core radius $\rm R_c$ and $\beta$ parametrize the gas
density profile, $\mu$ is the mean molecular
weight ($\mu$=0.6 for a primordial composition with a 76\% fraction
contributed by hydrogen) and $\rm m_{p}$ the proton mass. The mass
considered in the self--similar model, expressed as a function of the
radius $R_{\Delta_{z}}$ which encloses a mean density which is a 
factor $\Delta_z$ above the critical density of the Universe, is given
by: 
\begin{eqnarray} 
{\rm M_{tot} (< R_{\Delta_{z}})} = \frac{4 \pi}{3} {\rho}_{c,z} {
\Delta}_{z} {\rm R^3_{\Delta_{z}}} 
\label{eqn4}
\end{eqnarray}   
\noindent
and:
\begin{eqnarray} 
\Delta_{z} = \Delta(0)\Delta_{\rm v}(z)/\Delta_{\rm v}(0) 
\label{eqn5}
\end{eqnarray}
\smallskip\noindent
where $\Delta(0)$ is the density contrast at z=0.
\noindent
From Eq. \ref{eqn3} and  \ref{eqn4} the radius corresponding
to $\Delta_{z}$ is calculated as follows:
\begin{eqnarray} 
{\rm R}_{\Delta_{z}} = \sqrt{\frac {6 \beta k_{B} {\rm T(^{\circ} K)}}
{\mu m_p} \frac{1}{H_z^2 \Delta_{z}} - {\rm R_c^2}}
\label{eqn6}
\end{eqnarray}
\noindent
where $H_z$ is defined in Eq. \ref{eqn1}.  For $\Delta_{z} =
\Delta_{\rm v}(z)$ one obtains the virial radius $R_{\rm v}(z)$.  
These radii are $\rm \propto T^{0.5}$ and decrease with redshift in a 
way dependent on the assumed cosmological model.

\smallskip\noindent
At very high energies the ICM behaves as a fully ionized plasma and
thus it emits by thermal bremsstrahlung in the X-ray band.  The
emissivity for this thermal process at frequency $\nu$ scales as:
\begin{eqnarray} 
\epsilon_{\nu} \propto n_e n_i g(\nu,T) T^{-1/2} \rm e^{-h\nu/k_{B}T}
\label{eqn7}
\end{eqnarray} 
\noindent
Ignoring the temperature dependence of the Gaunt factor $g(\nu,T)$
\citep[see, e.g.,][]{Sp78}, and assuming a constant ratio of gas
density to total mass density, the scaling behavior of a cluster X-ray
bolometric luminosity can be written as:
\begin{eqnarray}
\rm L_{bol} \propto M \rho_{gas} T^{1/2}
\label{eqn8}
\end{eqnarray} 
\noindent
Then one can obtain the following scaling relations between the observed 
bolometric luminosity $\rm L_{bol}$ and temperature of the gas T and 
the total gravitating mass: 
\begin{eqnarray}
{\rm L_{\rm bol}  \propto T^2}  E_z \Delta_z^{1/2} ~~~~~~{\rm and}~~~~~~ 
{\rm L_{\rm bol}  \propto M_{tot}^{4/3}} E_z^{7/3} \Delta_z^{7/6}
\label{eqn9} 
\end{eqnarray}

\noindent   
The relation between the two X-ray observables, temperature and
luminosity, is a useful statistical tool that enables the study of the
physics of clusters and allows the verification of the self--similarity
model expectations.

%----------------------------- subsection ----------------------------
\subsection{Evolution with redshift of the $\rm L_{bol}-T$ relation}
\label{defevo}

When a high-redshift cluster sample is analyzed, it is important to
define what is meant by evolution, since different definitions
are found in the literature. The self--similar models
predict changes with redshift of the scaling laws through the factors
$E_z$ and $\Delta_{z}$. For instance, the $\rm L_{bol}-T$ relation
(Eq. \ref{eqn9}) contains the cosmological term $E_z
\Delta_{z}^{1/2}$.  One definition of evolution (hereafter referred to
as Ev$_1$) is any change with redshift in the scaling laws not
accounted for by the $E_z$ and $\Delta_{z}$ factors.  A second
definition of evolution (hereafter referred to as Ev$_2$) 
accounts for any change with respect to the local observed 
relations, including  $E_z$ and $\Delta_{z}$.

\medskip\noindent 
In the case of Ev$_1$ the approach to follow is to correct the X-ray
luminosity for the $E_z$ and $\Delta_{z}$ factors and then to introduce
a factor $(1+z)^{A_z}$ to account for any remaining effect with
redshift.  The $\rm L_{bol}-T$ relation then becomes:
$E_z^{-1}(\Delta_z/\Delta_{z=0})^{-1/2}{\rm L_{bol}} \propto 
{\rm T^{\alpha}} (1 + z)^{A_{z}}$. With this notation in the 
self--similar scenario $A_z$ = 0 and no Ev$_1$ is present.

\noindent 
For Ev$_2$ generally a factor $(1+z)^A$ is used which includes 
also the $E_z$ and $\Delta_{z}$ factors. For the $\rm L_{bol}-T$
relation one would write: ${\rm L_{bol} \propto T^{\alpha}} (1 +
z)^{A}$.  In an Einstein-de Sitter Universe the cosmological factor
$E_{z}$ (Eq.\ref{eqn1}) and the overdensity $\Delta_z$
(Eq.\ref{eqn5}) are equal to $(1+z)^{1.5}$ and 1, respectively.
The self--similar scaling relation would be: ${\rm L_{bol} = C
T^{\alpha}} (1+z)^{1.5}$, and therefore $A$ = 1.5.  We would talk of
$evolution$ according to the definition of Ev$_2$, but $no~ evolution$
according to the definition of Ev$_1$.

\noindent  
In a $\Lambda$CDM model with $\Omega_{m}=0.3$, the dependence of
$E_{z}\Delta_z^{1/2}$ is no longer a power law of (1+z) but it can be
approximated with that law in a limited range of redshifts by the
redshift dependent exponent A :
\begin{eqnarray}
A = \frac {\log \left[\left( \frac{E_{z}}{E_{z=0}}\right)  
{\left( \frac{\Delta_z}{\Delta_{z=0}} \right)}^{1/2}\right]}{{\log(1+z)}}, 
~~~{\rm where}~~E_{z=0} = 1
\label{eqn10}
\end{eqnarray}

\noindent
In the case of evolution Ev$_1$ according to the redshift interval one
can calculate the range of predictions for $A$ on the basis of the
self--similarity model and evaluate any possible evolution with respect
to self--similarity. For example for a redshift range 0.3$\div$1.0 the
self--similar model prediction for $A$ is in the range $1 \div 1.1$.

%----------------------------- subsection ----------------------------
\subsection{Breaking the Self--Similar Model}
\label{breakselfsim}

One way to break the scaling laws that predict $\rm L_{bol}\propto
T^2$ is to have non gravitational energy injected into the
intracluster medium before or during the cluster formation. 
An abundant literature exists on this subject 
\citep[for a review see e.g.][and reference therein]{Vo05a,Toz06}.
The candidates for the energy excess are currently feedback from star
formation processes and feedback from nuclear activity in the cluster
galaxies.  Another physical process that could break the simple
self--similar scaling is the removal of low-entropy gas via cooling
\citep[e.g.][]{Bry00,Vo01}.

\smallskip\noindent
For example, \cite{Vo05a} gives the following $\rm L_{bol}-T$ relations
(his equations 81, 82, 83) for three different models:

    $ {\rm L_{bol} \propto T^{3.5}} (E_z)^{-1}$
    
    $ {\rm L_{bol} \propto T^{2.5}} (E_z)^{-1} t(z)^{-1}$
    
    $ {\rm L_{bol} \propto T^{3}} (E_z)^{-3} t(z)^{-2}$
    
\noindent
where $t(z) $ is the cosmic time at redshift z. 
The above relations were obtained by using a cluster radius that encloses 
a mean density which is a fixed factor ($\Delta = 200$) with respect to the 
critical density.  This  alternative choice of a fixed overdensity is often found 
in the literature \citep[see e.g.][]{Ev02}.  
In this case the factor $\Delta_z$ is  constant and the redshift dependence 
in the self-similar scaling relations (Eq.s \ref{eqn9}) is only through the 
factor $E_z$.

\smallskip\noindent
The first model is a pure {\it pre-heating} case, in which the
minimum entropy is assumed to be independent of both cluster mass and
redshift.  The other two are cooling models. These models give a slope
of the $\rm L_{bol}-T$ relation significantly steeper than the
self--similar model.  They also predict different dependences on
redshift (approximately $\propto (1+z)^{-0.8}, (1+z)^{0.4}$, and
almost constant in the redshift range 0 -- 1).
The relations differ from the self--similar ones in their dependence
on T and on z. Therefore, in principle, any observed change in the
scaling laws allows to constrain different scenarios of heating
schemes and cooling efficiency.  

%----------------------------- Section ----------------------------
\section{The cluster sample}

In this work we use {\em Chandra} archival data of eighteen distant
(0.25 $<$ z $<$ 1.01) clusters. Details of the X--ray analysis are
given in a companion paper \citep [][Paper I]{Br07a}.  
This same sample was also used by \cite{Br07b} to check for the presence
of  overdensity of point sources in the inner region of clusters of 
galaxies.

\smallskip\noindent
Since the $\rm L_{bol}-T$ relationship at both low and high redshift
exhibits a large intrinsic scatter, mostly due to the strong cooling 
flow clusters \citep{Fa94}, the presence of cooling cores has 
to be considered. To reduce the intrinsic scatter, which precludes the 
accurate determination of the exact shape and of any possible evolution of 
the $\rm L_{bol}-T$ relation, one can exclude from the sample the clusters 
with strong cooling cores \citep[e.g.][]{Ar99,Et04,Ma06}.
Alternatively, since cooling flow clusters are similar to non--cooling 
flow clusters except for the inner small cooling flow region, an approach 
to follow is to excise the cluster central cores and replace them with an
extrapolation of the isothermal $\beta$-model before deriving cluster
parameters \citep[e.g.][]{Ma98,Vi02}. There are four clusters in our sample 
which \cite{Vi02} indicate as ``possible cooling core" clusters.  We have
kept these clusters in the sample without applying any correction to T
or to L$_{\rm bol}$ since according to their radial profile derived by
us these cooling cores, if present at all, are rather mild. A fifth
cluster, ZW\,CL\,1454.8$+$2233 presents a very thermally complex ICM 
\citep [see notes on individual clusters in Paper I,][]{Br07a}
and was thus excluded from the analysis. From now on the sample of 
17 clusters is referred to as `17 cluster sample'.

\smallskip\noindent 
In order to increase the statistics we supplemented our data with the
sample of \cite{Vi02} (from now on VI02), \cite{Et04} (from now on
ET04) and of \cite{Ma06} (form now on MA06) after appropriate
corrections.  For clusters in common we use our own measurements. For
clusters not in our sample but in common with the other three samples,
we preferred the data of MA06 when available, since they use the most
recent calibrations of {\em Chandra}. We also used some of the MA06
clusters which were observed with XMM-{\it Newton}.  For clusters
common to VI02 and ET04 samples, we use the VI02 clusters since their
data are in better agreement with ours after corrections for the
different radii are made \citep [see][Paper I]{Br07a}. This
choice was made ``a posteriori'' since the $\chi^2$ analysis
evaluation of the combined data, given in Sect. ~\ref{ltanaly}, shows
a worse value when using ET04 data, even if the final results do not
change independently of the sample used.  The final working sample is
thus composed by 17 clusters from our sample, 8 clusters from VI02, 10
clusters from MA06 and 4 clusters from ET04 for a total of 39
clusters. From now on the sample of 39 clusters is referred to as
`combined sample'.

\smallskip\noindent
Since we want to compare our observational results with theoretical
models, the choice of the radius within which integrated cluster
properties are measured becomes very important, notably when a large
range of redshits is involved. Several authors have chosen a constant
radius which has the great advantage of simplicity. A constant radius
would allow the analysis of the same fractional volume of clusters if
clusters have all the same physical radius. However, in the 
self--similar models the virial radius is a function of redshift and
temperature, thus a constant physical radius does not include a constant
fraction of the cluster, and  introduces a bias with redshift in the
computed cluster luminosities. Since the self--similar models predict
the outer boundary of the virialized part of clusters in terms of a
density contrast $\Delta_{\rm v}(z)$ \citep[see for example][]{Bry98}, we
consider appropriate to calculate the sample luminosities using the
radius where the mean enclosed density is a factor $\Delta_z$
above the critical density of the Universe \citep{Et04,Ma06}.  In this
way a fixed fraction of virial radius is used, independent of the
cluster redshift, with no redshift and temperature biases in the
luminosities. This choice allows us to use in a straightforward way
the dependence of cluster properties on redshift as reviewed
earlier (see Sect.~\ref{selfsim}, \ref{breakselfsim}). In the present
analysis we use for the overdensity $\Delta_{z} = 500 \times
\Delta_{\rm v}(z)/(18\pi^2)$ as in ET04. Note that the 
$\Delta_{z=0}$ value is $\approx 280$, while the overdensity in a 
flat $\Lambda$ = 0 cosmology would be $500$. For each cluster the
value of $\rm R_{500}$ is calculated according to Eq. \ref{eqn6}.
All luminosities are extrapolated to $\rm R_{500}$ using an
isothermal $\beta$-profile.

\smallskip\noindent
The uncertainties on $\rm R_{500}$ and $\rm L_{500}$ are calculated
using the propagation of the measurement uncertainties on temperature
and luminosities (which are the dominant contribution) and of the
uncertainties related to the $\beta$-model.  
Since core radius, $\rm R_c$, and $\beta$ are correlated quantities
(surface-brightness model with large values of $\rm R_c$ and $\beta$ 
are similar to those with small values), the uncertainties on them
are not independent. Thus the errors on $\rm R_c$ were ignored
in the calculation of the uncertainties (see MA06).

\smallskip\noindent
It has to be noted that there may be some inconsistencies when
comparing the high-z combined sample with the local samples due to a
different choice of the radius used for the luminosity estimates.
\cite{Ma98} adopted a radius of 1.4 Mpc, independent of
temperature and $\beta$, which is however close to $R_{500}$ for
average temperatures and $\beta$ values due to the low redshift of
their clusters. An evaluation of the effects due to different choices
of the radius resulted in negligible differences as compared to the errors.
\cite{Ar99} do not specify any radius. We guess that they used a value
encompassing all the visible X--ray emission. Even in this case we 
expect that systematic effects are negligible.

\smallskip\noindent
The temperature and the $\rm L_{500}$ luminosities for all the clusters 
used to analyze the $\rm L_{bol}-T$ relation are listed in Table~\ref{tab1}.
It is worth to note that no dependence on redshift is present for either
the luminosity or temperature of the sample clusters.
The columns in Table~\ref{tab1} contain the following information:
 \begin{itemize}
 \item [--] Column 1: Cluster name. Asterisks indicate ``possible cooling 
            core''  clusters as classified by VI02. However, no correction was 
            applied to the four objects in the `17 cluster sample' since we checked 
            that the cooling core, if any, is negligible.
 \item [--] Column 2: Spectroscopic redshift
 \item [--] Column 3: Cluster temperature in keV. For VI02, 
            ET04 and MA06 the temperature is taken from their papers.  
 \item [--] Column 4-5: Core radius in kpc and $\beta $. For VI02, ET04 
            and MA06 these parameters are taken from their papers. 
 \item [--] Column 6: Radius in kpc which corresponds to a redshift-dependent 
            contrast $\Delta_{z} = 500~\times~\Delta_{\rm v}(z)/(18\pi^2)$.
            For each cluster the radius $\rm R_{500}$ is computed according 
            to Eq. \ref{eqn6}.
 \item [--] Column 7: Luminosity within $\rm R_{500}$ expressed in 
            $\rm 10^{44} erg\ s^{-1}$. For the `17 cluster sample' the 
            estimated luminosities \citep [see][Paper I]{Br07a}
            are extrapolated to $\rm R_{500}$ using a 
            $\beta$-model. In the same way the VI02 luminosities (estimated 
            within the fixed radius of 1.4 h$^{-1}_{70}$ Mpc) and the MA06 
            luminosities (estimated within the radius $\rm R_{200}$ 
            corresponding to an overdensity  $\Delta_{z} = 
            200~\times~\Delta_{\rm v}(z)/\Delta_{\rm v}(0)$) were extrapolated 
            to $\rm R_{500}$
 \item [--] Column 8: Original sample 
\end{itemize}
%-------------------Table 1--------------------------------------------------
\begin{table}[htb]
\begin{center}
\caption{Combined sample: cluster parameters}
\begin{tabular}{lccrrrrc}
\hline
\\
Cluster name & z & T & $\rm R_{c}$   &  $\beta$ &  $\rm R_{500}$ & $\rm L_{500}$ & Sample \\
\hline
\hline
Abell\,2125               & 0.246 &  3.4$_{~-~0.2}^{~+~0.2}$ &182$\pm$12 & 0.54$\pm$0.02& 888$\pm$28 &   2.13$\pm$0.08& `17 cluster' \\    
MS\,1008.1-1224           & 0.302 &  6.0$_{~-~0.3}^{~+~0.4}$ &165$\pm$9  & 0.64$\pm$0.02&1241$\pm$42 &  11.89$\pm$0.37& 	    \\
ZW\,CL\,0024.0+1652$^{*}$ & 0.394 &  4.4$_{~-~0.4}^{~+~0.5}$ &128$\pm$10 & 0.67$\pm$0.03&1003$\pm$58 &   4.38$\pm$0.25& 	    \\
MS\,1621.5+2640           & 0.426 &  7.5$_{~-~0.7}^{~+~1.1}$ &227$\pm$17 & 0.65$\pm$0.03&1255$\pm$84 &  12.98$\pm$0.69& 	    \\
RXJ\,1701.3+6414$^{*}$    & 0.453 &  4.5$_{~-~0.3}^{~+~0.4}$ & 15$\pm$2  & 0.41$\pm$0.01& 765$\pm$33 &   7.14$\pm$0.40& 	    \\
CL\,1641+4001             & 0.464 &  5.1$_{~-~0.7}^{~+~0.8}$ &151$\pm$18 & 0.77$\pm$0.06&1092$\pm$91 &   3.02$\pm$0.25& 	    \\
V\,1524.6+0957            & 0.516 &  5.0$_{~-~0.5}^{~+~0.6}$ &302$\pm$27 & 0.80$\pm$0.05&1029$\pm$72 &   7.10$\pm$0.52& 	    \\
MS\,0451.6-0305           & 0.539 &  9.4$_{~-~0.5}^{~+~0.7}$ &270$\pm$8  & 0.90$\pm$0.02&1505$\pm$51 &  52.89$\pm$2.75& 	    \\
V\,1121+2327              & 0.562 &  4.5$_{~-~0.4}^{~+~0.5}$ &437$\pm$58 & 1.19$\pm$0.18&1108$\pm$116&   5.73$\pm$0.43& 	    \\
MS\,2053.7-0449$^{*}$     & 0.583 &  4.3$_{~-~0.4}^{~+~0.5}$ &115$\pm$12 & 0.64$\pm$0.03& 839$\pm$49 &   5.81$\pm$0.55& 	    \\
V\,1221+4918              & 0.700 &  7.0$_{~-~0.7}^{~+~0.8}$ &272$\pm$20 & 0.76$\pm$0.04&1037$\pm$64 &  13.22$\pm$0.79& 	    \\
MS\,1137.5+6625           & 0.782 &  6.2$_{~-~0.4}^{~+~0.5}$ &116$\pm$6  & 0.71$\pm$0.02& 910$\pm$39 &  13.95$\pm$0.53& 	    \\
RDCSJ\,1317+2911$^{*}$    & 0.805 &  3.7$_{~-~0.8}^{~+~1.2}$ & 61$\pm$16 & 0.52$\pm$0.04& 592$\pm$87 &   1.35$\pm$0.40& 	    \\
RDCSJ\,1350+6007          & 0.805 &  4.1$_{~-~0.6}^{~+~0.8}$ &261$\pm$43 & 0.70$\pm$0.07& 688$\pm$77 &   5.33$\pm$0.78& 	    \\
RXJ\,1716.4+6708          & 0.813 &  6.5$_{~-~0.8}^{~+~0.9}$ &119$\pm$11 & 0.66$\pm$0.03& 882$\pm$61 &  13.35$\pm$1.08& 	    \\
MS\,1054.4-0321           & 0.830 &  8.3$_{~-~0.7}^{~+~0.7}$ &520$\pm$32 & 1.38$\pm$0.11&1339$\pm$89 &  35.73$\pm$1.79& 	    \\
WARPJ\,1415.1+3612        & 1.013 &  6.2$_{~-~0.7}^{~+~0.8}$ & 68$\pm$7  & 0.60$\pm$0.02& 722$\pm$43 &  12.27$\pm$0.88& 	    \\
\hline			            			 	  								    
CL\,1416+4446$^{*}$       & 0.400 &  3.7$_{~-~0.3}^{~+~0.3}$ &152	 & 0.66$\pm$0.04& 906$\pm$47 &   4.14$\pm$0.29&       VI02  \\
MS\,0302+1658$^{*}$       & 0.424 &  3.6$_{~-~0.5}^{~+~0.5}$ &154	 & 0.74$\pm$0.09& 929$\pm$88 &   5.15$\pm$0.37& 	    \\
RASS\,1347-114$^{*}$      & 0.451 & 14.1$_{~-~0.9}^{~+~0.9}$ &84	 & 0.65$\pm$0.01&1706$\pm$56 & 135.01$\pm$9.45& 	    \\
3C\,295$^{*}$	          & 0.460 &  5.3$_{~-~0.5}^{~+~0.5}$ &103	 & 0.67$\pm$0.03&1051$\pm$55 &   8.04$\pm$0.56& 	    \\
MS\,0016+1609             & 0.541 &  9.9$_{~-~0.5}^{~+~0.5}$ &274	 & 0.74$\pm$0.02&1396$\pm$41 &  57.38$\pm$4.03& 	    \\
CL\,0848+4456             & 0.574 &  2.7$_{~-~0.3}^{~+~0.3}$ & 78	 & 0.58$\pm$0.04& 636$\pm$42 &   1.29$\pm$0.09& 	    \\
RDCS\,0910+5422           & 1.100 &  3.5$_{~-~0.7}^{~+~0.7}$ &111	 & 0.72$\pm$0.18& 550$\pm$92 &   2.66$\pm$0.21& 	    \\
RDCS\,0849+4452           & 1.261 &  4.7$_{~-~1.0}^{~+~1.0}$ &119	 & 0.85$\pm$0.33& 628$\pm$144&   2.95$\pm$0.22& 	    \\
\hline			            			 	  								    
RXJ\,0542-4100            & 0.634 &  7.9$_{~-~0.8}^{~+~1.0}$ &132$\pm$17 & 0.51$\pm$0.03& 982$\pm$63 &  12.15$\pm$1.36&       ET04  \\
RXJ\,2302+0844            & 0.734 &  6.6$_{~-~0.6}^{~+~1.5}$ & 96$\pm$12 & 0.55$\pm$0.03& 865$\pm$65 &   5.45$\pm$0.17& 	    \\
RXJ\,1252-2927            & 1.235 &  5.2$_{~-~0.7}^{~+~0.7}$ & 77$\pm$13 & 0.53$\pm$0.03& 532$\pm$40 &   5.99$\pm$1.10& 	    \\
RXJ\,0848+4453            & 1.270 &  2.9$_{~-~0.8}^{~+~0.8}$ &163$\pm$70 & 0.97$\pm$0.43& 499$\pm$115&   1.04$\pm$0.73& 	    \\
\hline		                    			 	  								    
ClJ\,0046.3+8530          & 0.62  &  4.4$_{~-~0.4}^{~+~0.5}$ &137$\pm$28 & 0.60$\pm$0.06& 792$\pm$56 &   3.82$\pm$0.22&       MA06  \\ 
ClJ\,1342.9+2828          & 0.71  &  3.7$_{~-~0.4}^{~+~0.5}$ &172$\pm$26 & 0.70$\pm$0.06& 724$\pm$55 &   3.34$\pm$0.18& 	    \\
ClJ\,1113.1-2615          & 0.73  &  4.7$_{~-~0.7}^{~+~0.9}$ &106$\pm$12 & 0.67$\pm$0.04& 801$\pm$74 &   3.73$\pm$0.32& 	    \\
ClJ\,1103.6+3555          & 0.78  &  6.0$_{~-~0.7}^{~+~0.9}$ &141$\pm$18 & 0.58$\pm$0.11& 807$\pm$93 &   4.62$\pm$0.30& 	    \\
ClJ\,0152.7-1357N         & 0.83  &  5.6$_{~-~0.8}^{~+~1.0}$ &249$\pm$48 & 0.73$\pm$0.10& 820$\pm$93 &  10.68$\pm$1.42& 	    \\
ClJ\,0152.7-1357S         & 0.83  &  4.8$_{~-~1.0}^{~+~1.1}$ &123$\pm$24 & 0.66$\pm$0.07& 744$\pm$93 &   6.42$\pm$0.96& 	    \\
ClJ\,1559.1+6353          & 0.85  &  4.1$_{~-~1.0}^{~+~1.4}$ &67 $\pm$30 & 0.59$\pm$0.09& 646$\pm$107&   2.47$\pm$0.31& 	    \\
ClJ\,1008.7+5342          & 0.87  &  3.6$_{~-~0.6}^{~+~0.8}$ &170$\pm$43 & 0.68$\pm$0.09& 622$\pm$79 &   3.57$\pm$0.35& 	    \\
ClJ\,1226.9+3332          & 0.89  & 10.6$_{~-~1.1}^{~+~1.1}$ &113$\pm$8  & 0.66$\pm$0.02&1070$\pm$58 &  42.51$\pm$1.00& 	    \\
ClJ\,1429.0+4241          & 0.92  &  6.2$_{~-~1.0}^{~+~1.5}$ & 97$\pm$9  & 0.67$\pm$0.06& 806$\pm$90 &   9.32$\pm$0.86& 	    \\
\\
\hline
\label{tab1}
\end{tabular}
\end{center}
\vspace{-1.2cm}
\end{table}
%------------------------------------------------------------------------------

\section{The L$_{\rm bol}$--T relationship for high-z clusters: the
analysis method}
\label{ltanaly}

In order to examine how cluster luminosity scales with
temperature and whether the $\rm L_{bol}$--T relation depends on
redshift, we followed four different steps in our analysis.

\smallskip\noindent
We first assumed the following evolution law (Ev$_2$):
\begin{eqnarray}
{\rm L_{bol,44}} = C {\rm T_{6}^{\alpha}} (1+z)^A
\label{eqn11}
\end{eqnarray}
\noindent
where $\rm L_{bol,44}$ is the bolometric luminosity in units of $10^{44}$ erg 
~s$^{-1}$ and $\rm T_6 = T(keV)/6$.
\noindent 
The parameters C, $\alpha$ and A are determined by a $\chi^2$ fit
which takes into account the uncertainties on the luminosity and 
temperature ($\sigma_{L}$, $\sigma_{T}$). The $\chi^2$ is written as:

\begin{eqnarray}
\chi^{2} = \sum_{\rm i=1}^{N} \frac{{[\log {\rm L_{bol,44,i} - log C 
- \alpha log T_{6,i}} - A  \log(1 + z_{\rm i}) ]}^2}{\epsilon^2_{\rm L_i} + \alpha^2 \epsilon^2_{\rm T_i}} 	
\label{eqn12}
\end{eqnarray}
where N is the number of clusters, $\rm \epsilon_T$ is equal to $\rm
\sigma_{T_{6}}/(T_6 \ln 10)$ and $\rm \epsilon_L$ is equal to $\rm
\sigma_{L}/(L \ln 10)$. The denominator represents the propagation of
the uncertainties on the measured quantities.  The minimum $\chi^{2}$
is found by varying $\log~C$, $A$ and $\alpha$ in a grid of values.

\begin{description}
\item[{\it a})] The fit is performed setting $A$ equal to
      zero. In this way we analyze the $\rm L_{bol}-T$ relation
      independently of the cluster redshift.
\item[{\it b})] The three parameters $A$, $\alpha$ and $\log~C$ are used 
      as free parameters in the fit.
\item[{\it c})] In order to constrain the evolution in the $\rm L_{bol}-T$
     relation with respect to nearby galaxy clusters, an approach similar
     to VI02 and ET04 is adopted. The fit is  performed by fixing in
     Eq. \ref{eqn12} $\alpha$ and log C to the best fit values 
     ({$\overline \alpha$}, log $\overline{C}$) obtained from two local 
     samples, namely those by \cite{Ma98} and \cite{Ar99}.
     \cite{Ma98} analyzes 30 nearby clusters with measured
     \textit{ASCA} temperatures and {\em ROSAT} luminosities. His
     cluster luminosities are computed within a constant radius (1.4
     Mpc for $H_0=70$ Km s$^{-1}$ Mpc$^{-1}$). \cite{Ar99} analyze 24
     nearby clusters with X-ray measurements from \textit{GINGA},
     \textit{ASCA} and \textit{Einstein}. Both samples have a median redshift 
     of 0.05.  In our cosmology the $\rm L_{bol}-T$ slope and normalization 
     found by \cite{Ma98} and \cite{Ar99} convert respectively to:
\begin{eqnarray}
\nonumber
\rm \overline \alpha=2.64 \pm 0.27{\rm~~and~~\log~\overline C}~=~0.80 \pm 0.04\\
\rm \overline \alpha=2.88 \pm 0.15{\rm~~and~~\log~\overline C}~=~0.77 \pm 0.03
\label{eqn13}
\end{eqnarray}
 \end{description}

\medskip\noindent
As a last step the luminosities are scaled by the cosmological factor
$E_z^{-1}(\Delta_z/\Delta_{z=0})^{-1/2}$ as predicted by the
self--similar model (see e.g. ET04 and MA06). The data are fitted
using the following evolution law (Ev$_1$):
\begin{eqnarray}
E_z^{-1}(\Delta_z/\Delta_{z=0})^{-1/2}{\rm L_{bol,44}} = C {\rm T_{6}^{\alpha}} 
(1+z)^{A_z}
\end{eqnarray}
\begin{description}
\item[{\it d})] The minimum $\chi^2$ is found either for $A_z$
      equal to zero and leaving $A_z$ , $\alpha$ and $\log~C$ as
      free parameters.
\end{description}
With this approach the redshift dependence expected by the
self--similar scenario is removed `a priori' and thus normalization and
slope values consistent with those from the local relations should be
obtained if the self--similar predictions on the evolution are
correct.

%----------------------------- section ----------------------------
\section{The L$_{\rm bol}$--T relationship for high-z clusters: results}
\subsection{Comparison with local samples}
\label{results}

\begin{figure}
\begin{center}
\resizebox{\textwidth}{!}{\includegraphics{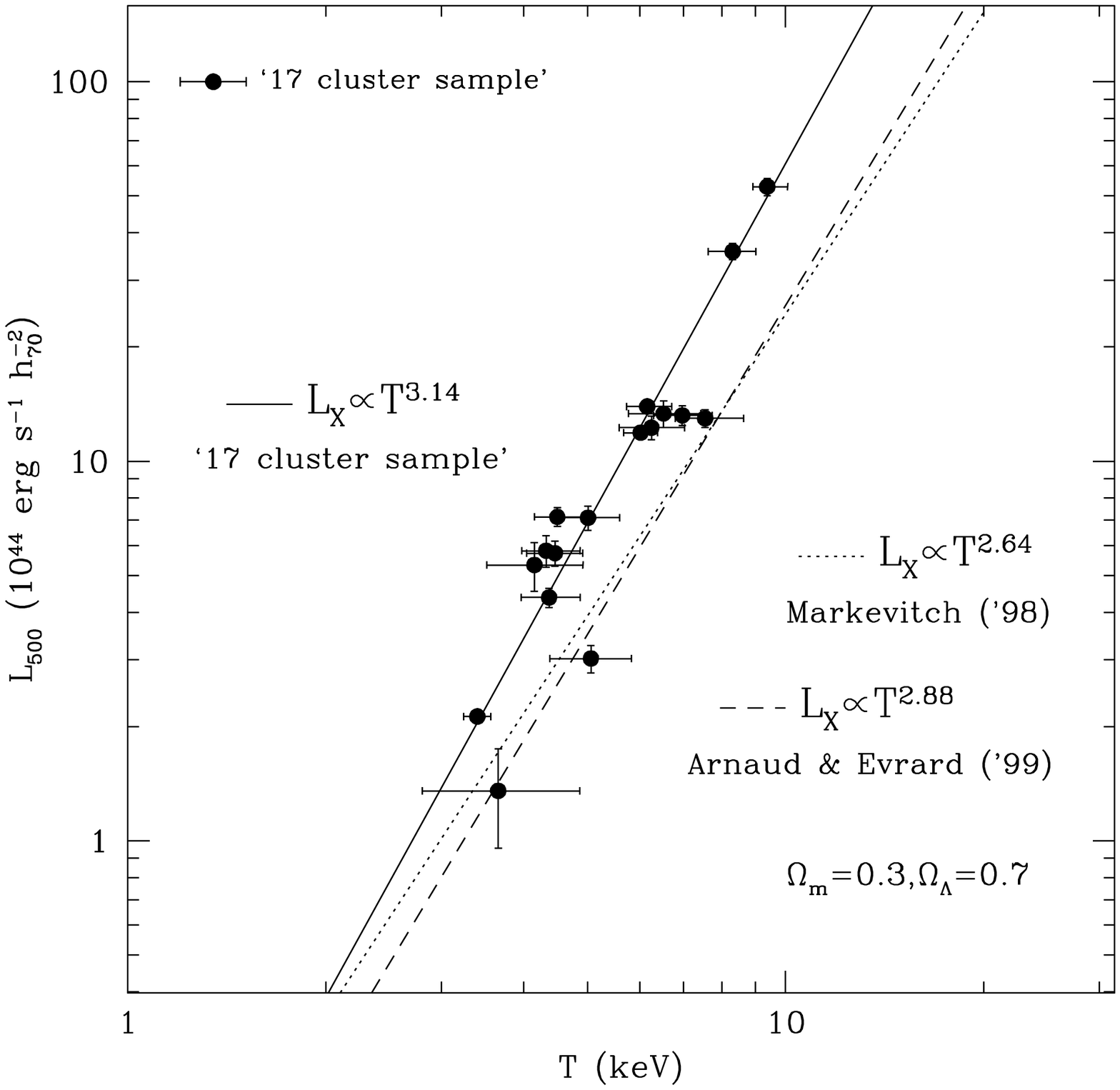}\includegraphics{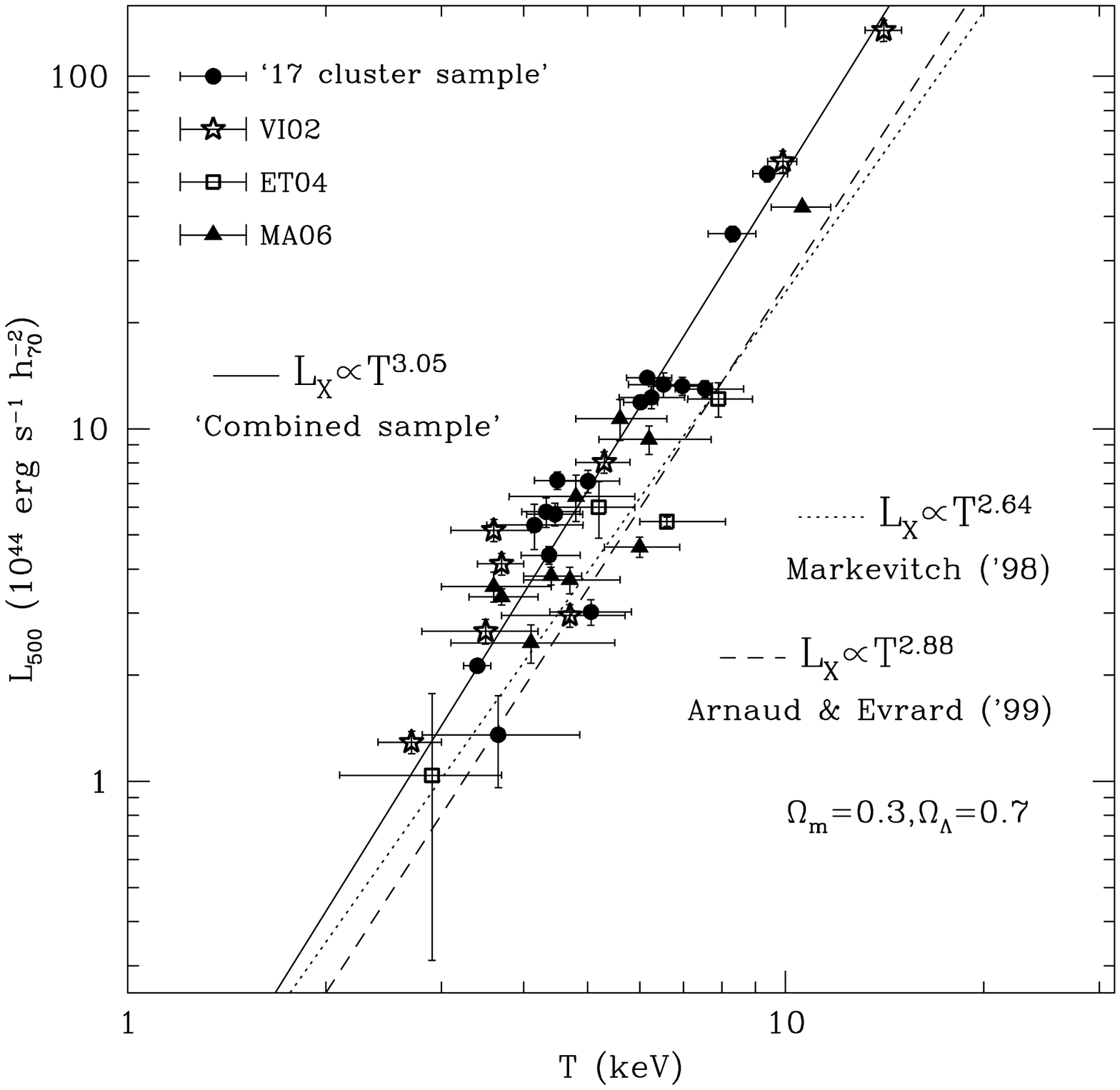}}
\end{center}
\caption{({\it Left panel}) The $\rm L_{bol}-T$ relation for the
sample of 17 distant clusters. The solid line indicates the best fit to
the data with $A$ set to 0. Dotted and dashed lines refer to the nearby 
cluster samples.
({\it Right panel}) Same as above but for the `combined 
sample' of 39 clusters which includes data from MA06, VI02 and ET04 as 
explained in the text. }
\label{fig1}
\end{figure}

The left panel of Fig.~\ref{fig1} shows the $\rm L_{bol}-T$ relationship for
the `17 cluster sample' while the right panel shows the same plot for
the `combined sample' of 39 clusters.  Overplotted are the best-fit
relationships found for the nearby clusters by \cite{Ma98} (dotted line) and
\cite{Ar99} (dashed line).
%--------------------------------- Figure 1------------------------------
%---------------------------------------------------------
A displacement of the distant $\rm L_{bol}-T$ relation from the local ones
is clearly evident, in agreement with the results of VI02, ET04 and MA06. 
High redshift clusters are on average systematically more luminous than the 
local ones by a factor $\approx$ 2.

\smallskip\noindent 
To quantify the differences between the high and low-z clusters the
$\rm L_{bol}-T$ fitting procedure, described in steps {\it a}) and
{\it b}), has been applied to the `17 cluster sample' and to the 
`combined sample'. 
The results of all the fits are listed in Table~\ref{tab2}.  
Setting A=0 (first and third line in Table~\ref{tab2}), the best-fit 
values for the slope $\alpha$ are somewhat higher but still consistent 
within the errors with the local ones (see Eq.s \ref{eqn13}) 
and significantly steeper than the self--similar predictions
($\alpha=2$). The best-fit values for $\log~C$ differ from the local
ones at more than 5 $\sigma$.

\smallskip\noindent 
When the Ev$_2$ evolution factor $(1+z)^A$ is introduced in the fit
(second and fourth line in Table~\ref{tab2}), the $\chi^2$
slightly improves but $\alpha$ and C do not change significantly
suggesting that the dependence on redshift must be weak. In fact the
Ev$_2$ parameter $A$ does not differ significantly from zero. Therefore
there is no evidence for a change in the normalization of the $\rm
L_{bol}-T$ relation as a function of redshift in the range covered by
the `combined sample' ($0.25 \le$ z $\le 1.3$) or, in other terms, no
Ev$_2$ is present.

\medskip\noindent 
The data were then fitted (step {\it c)}) by fixing
the slope and normalization to the best-fit relations obtained for the
low-z cluster samples of \cite{Ma98} and \cite{Ar99}. The results are
summarized in Table~\ref{tab3}. A significant positive evolution (of
Ev$_2$ type) with respect to the local relations is required: $A$
varies between 1.2 and 1.35 for the `combined sample'. A similar
positive evolution is found by other authors (e.g. VI02, Kotov \&
Vikhlinin 2005, Lumb et al. 2004, and MA06)\footnote{VI02 and
\cite{Ko05} found $A=$1.5 and $A=$1.8, respectively, somewhat larger
than our value. This is largely accounted for by their different
choice for the radius (1.4 Mpc) within which $\rm L_{bol}$ is
computed. Using their same radius and comparing our sample with
\cite{Ma98}, we find $A=$1.6 in good agreement with them.}. However,
the best-fits are significantly worsened with respect to the
previous step {\it b)} analysis.  In particular the $\chi^2$ of the
fit for the `combined sample' has a probability lower than 0.1\% to
be acceptable.  This evolution, which would be consistent with the
self--similar model predictions {($A \approx$ 0.9 $\div$ 1.2 for the
redshift range 0.25 $\div$ 1.3 with respect to redshift 0)}, is not
found in the high-z sample, but only when the entire redshift range
($0 \le$ z $\le 1.3$) is considered, that is using the local samples
as well.

\begin{figure}
\begin{center}
\resizebox{\textwidth}{!}{\includegraphics{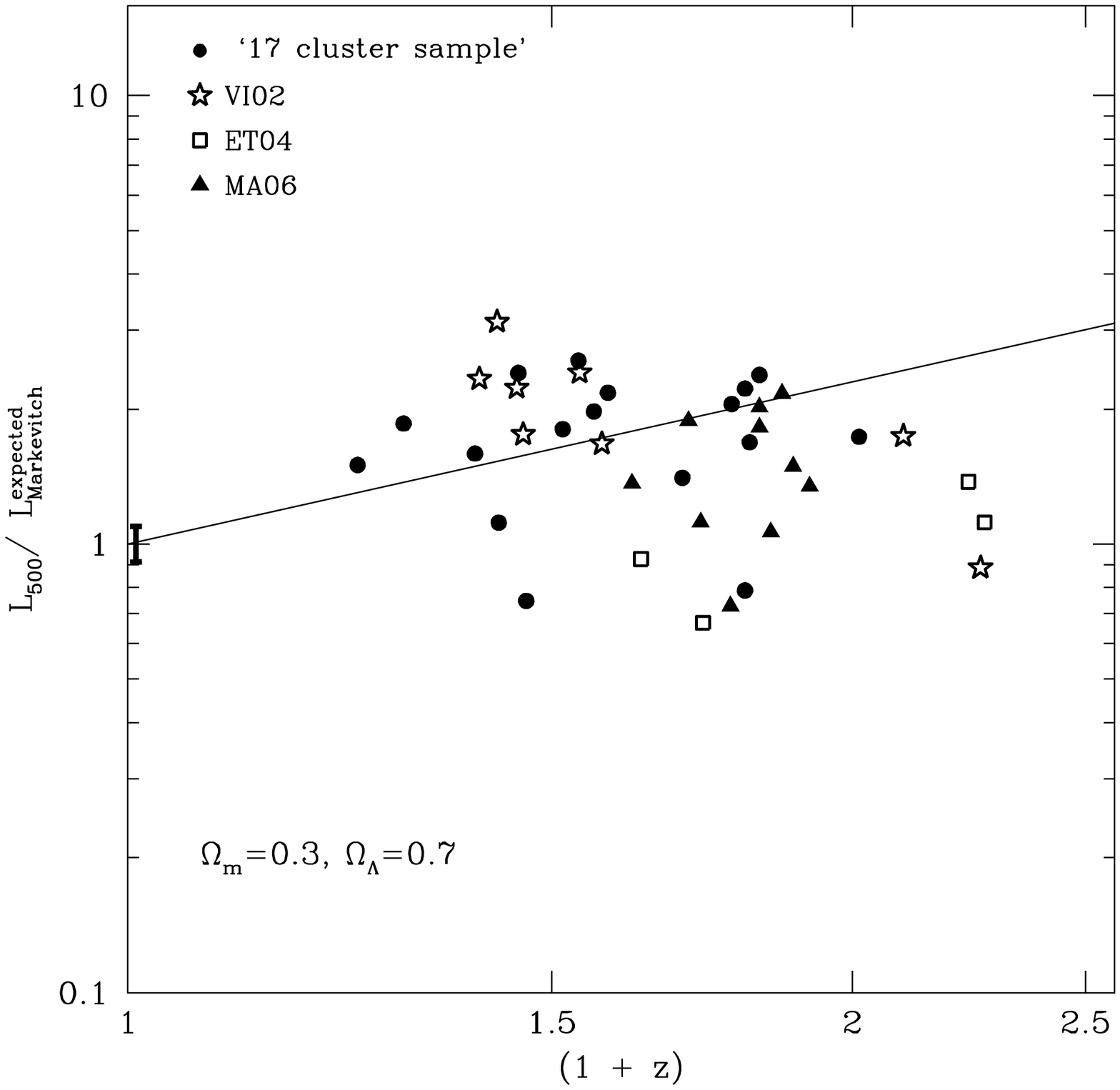}\includegraphics{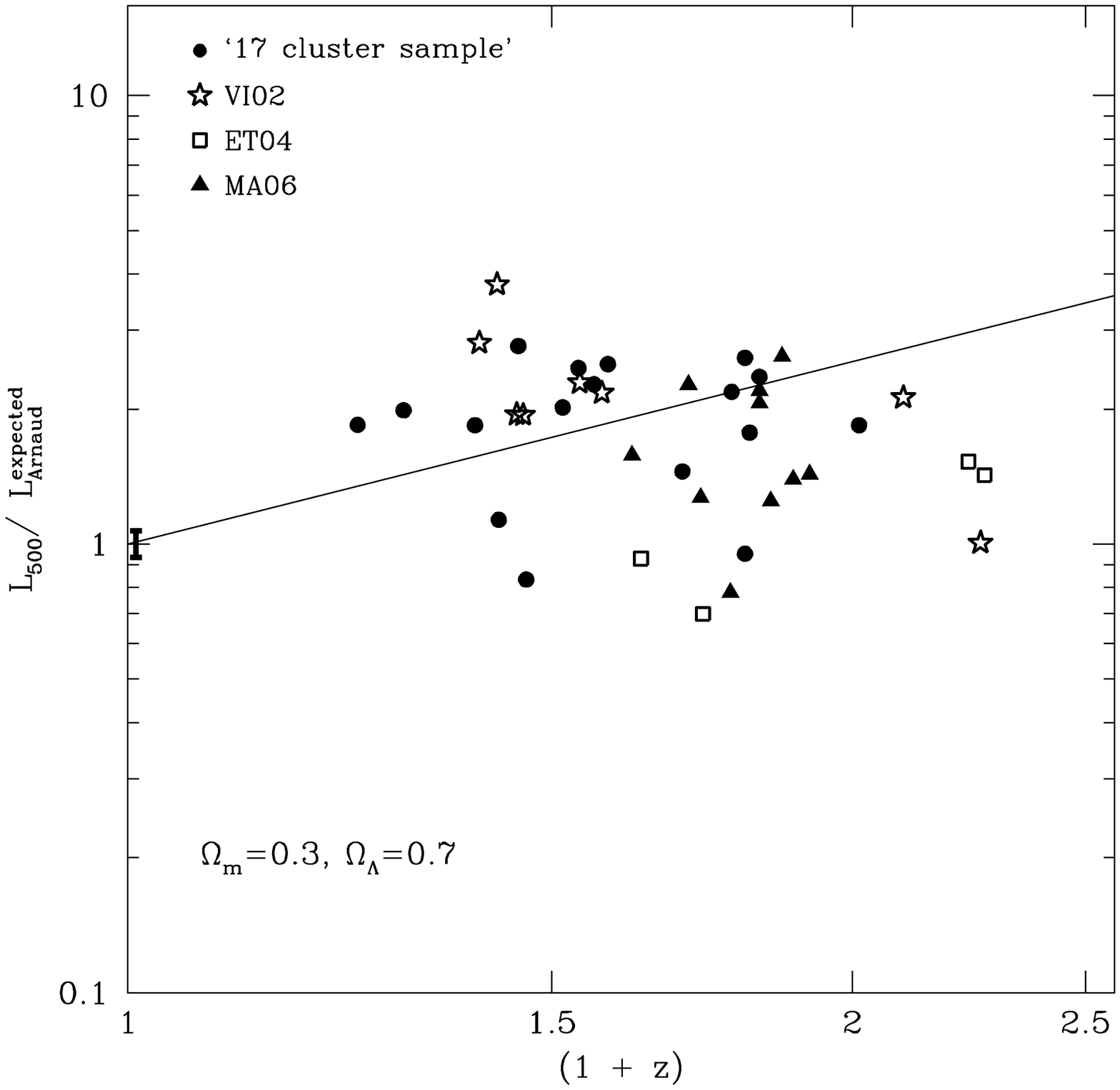}}
\end{center}
\caption{
Ratio of the observed luminosity to the `expected luminosity' from the
local relations (see Sect.~\ref{results}) versus $(1+z)$ for the
`combined sample'. ({\it Left panel}) The expected luminosity ($\rm
L^{expected}_{Markevitch}$) is the luminosity obtained from the local
relation of \cite{Ma98} using the observed cluster
temperature.  ({\it Right panel}) Same as above but for the local
relation of \cite{Ar99}.  In both panels the solid line indicates 
the fitted evolution law according to step {\it c)} of the
Sect.~\ref{ltanaly}. The evolution laws are clearly not
appropriate. The error bar at z=0 indicates the uncertainty on the
normalization of the local sample best-fit law.}
\label{fig2}
\end{figure}

\smallskip\noindent
In order to emphasize this result, we present in Fig.~\ref{fig2} 
the ratios between the observed cluster luminosity and the `expected 
luminosity' derived using the observed high--z cluster temperature in the
local relations by \cite{Ma98} and \cite{Ar99} as a function of (1+z). 
For each cluster of the `combined sample' the `expected luminosity' 
is calculated as:
\begin{eqnarray}
{\rm L^{expected}_{Markevitch,Arnaud}} = {\overline C} {\rm T_{6}^{\overline \alpha}} 
\end{eqnarray} 
where $\alpha$ and C are fixed to the best-fit values obtained by \cite{Ma98}
and \cite{Ar99} (see Eq.s~\ref{eqn13}), respectively, and $\rm T$ is the 
measure of the high--z cluster temperature.  Clearly the evolution law fitted according 
to  step {\it c)} (see Table~\ref{tab3}) and indicated by the solid line is not 
a good representation of the data.

%----------------------------Figure 2-----------------------------
%------------------------------------------------------------------------
%------------------------------------  Figure 3 --------------------------
\begin{figure}
\begin{center}
\resizebox{\textwidth}{!}{\includegraphics{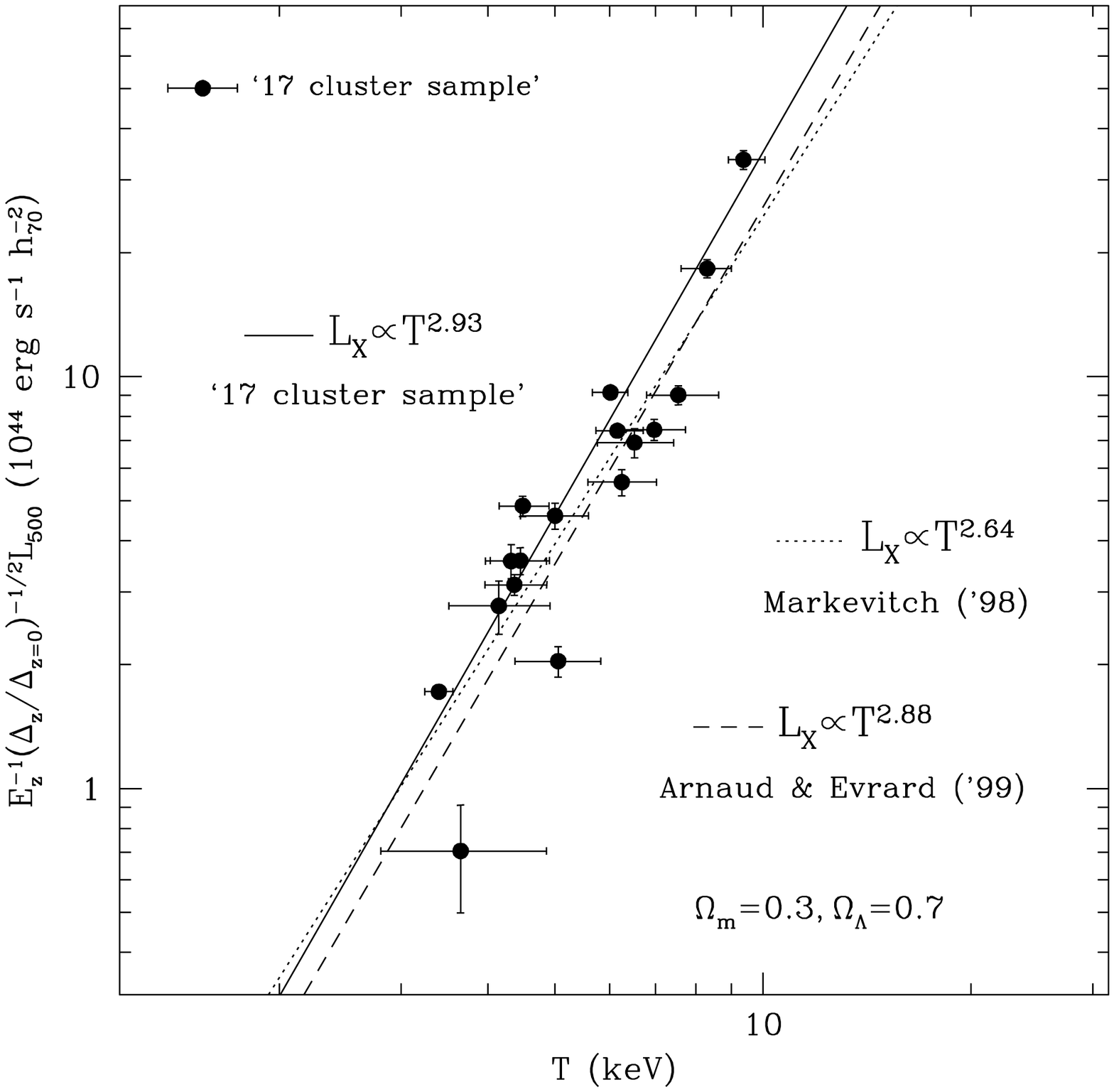}\includegraphics{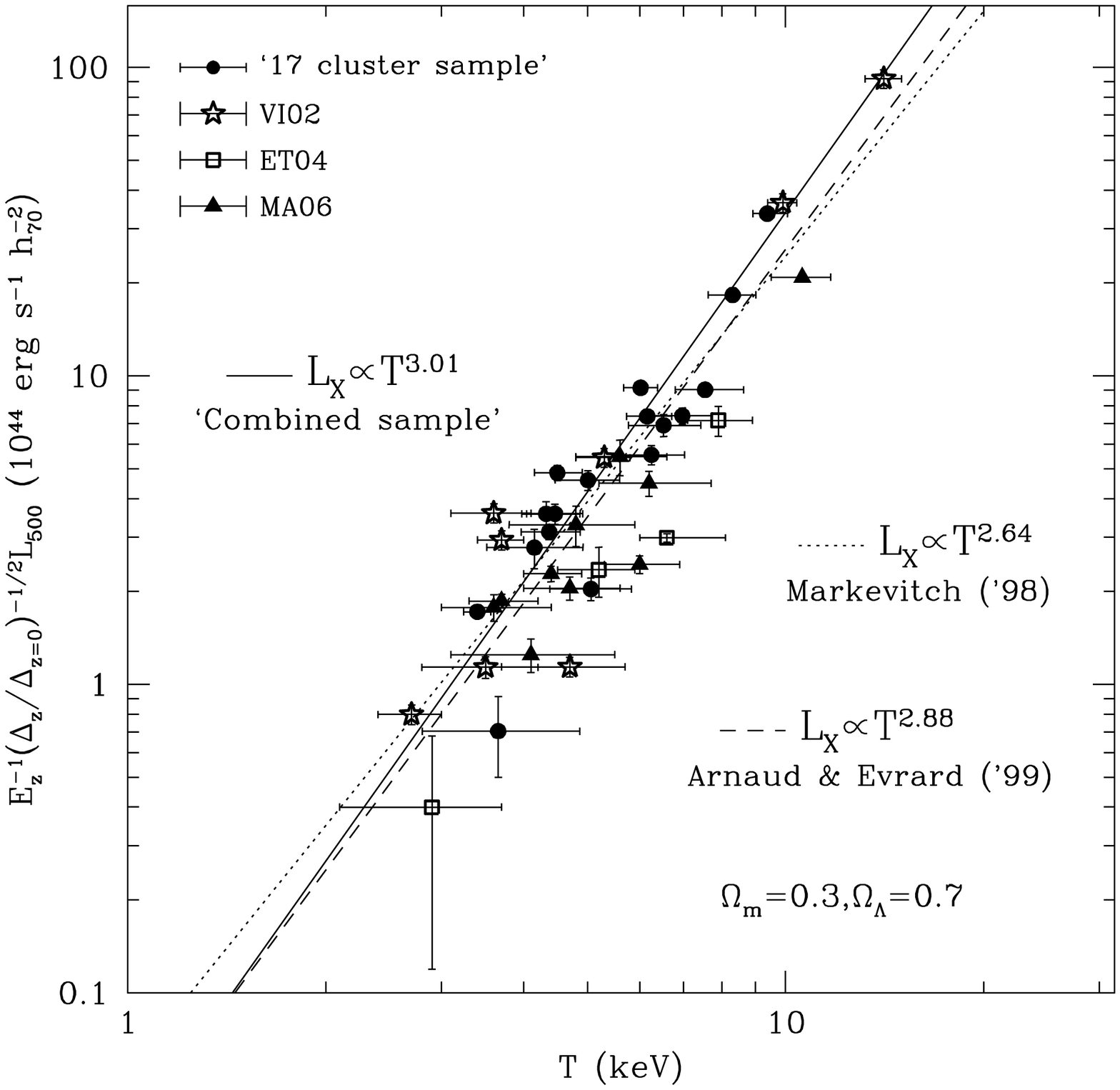}}
\end{center} 
\caption{ 
The $E_z^{-1} \left(\Delta_z/\Delta_{z=0}\right)^{-1/2}
{\rm L_{bol}-T}$ relation for the `17 cluster sample' ({\it left panel}) and
for the combined 39 cluster sample ({\it right panel}). The luminosity of
each cluster is scaled by the cosmological factor $E_z^{-1}
\left(\Delta_z/\Delta_{z=0}\right)^{-1/2}$. The solid line indicates
the best-fit to the data obtained fixing $A_{z}=0$.  The dotted and  
dashed lines indicate the local relations.}
\label{fig3}
\end{figure} 
%------------------------------------------------------------------------

\subsection{Comparison with the self--similar model}

\begin{figure}
\begin{center}
\resizebox{\textwidth}{!}{\includegraphics{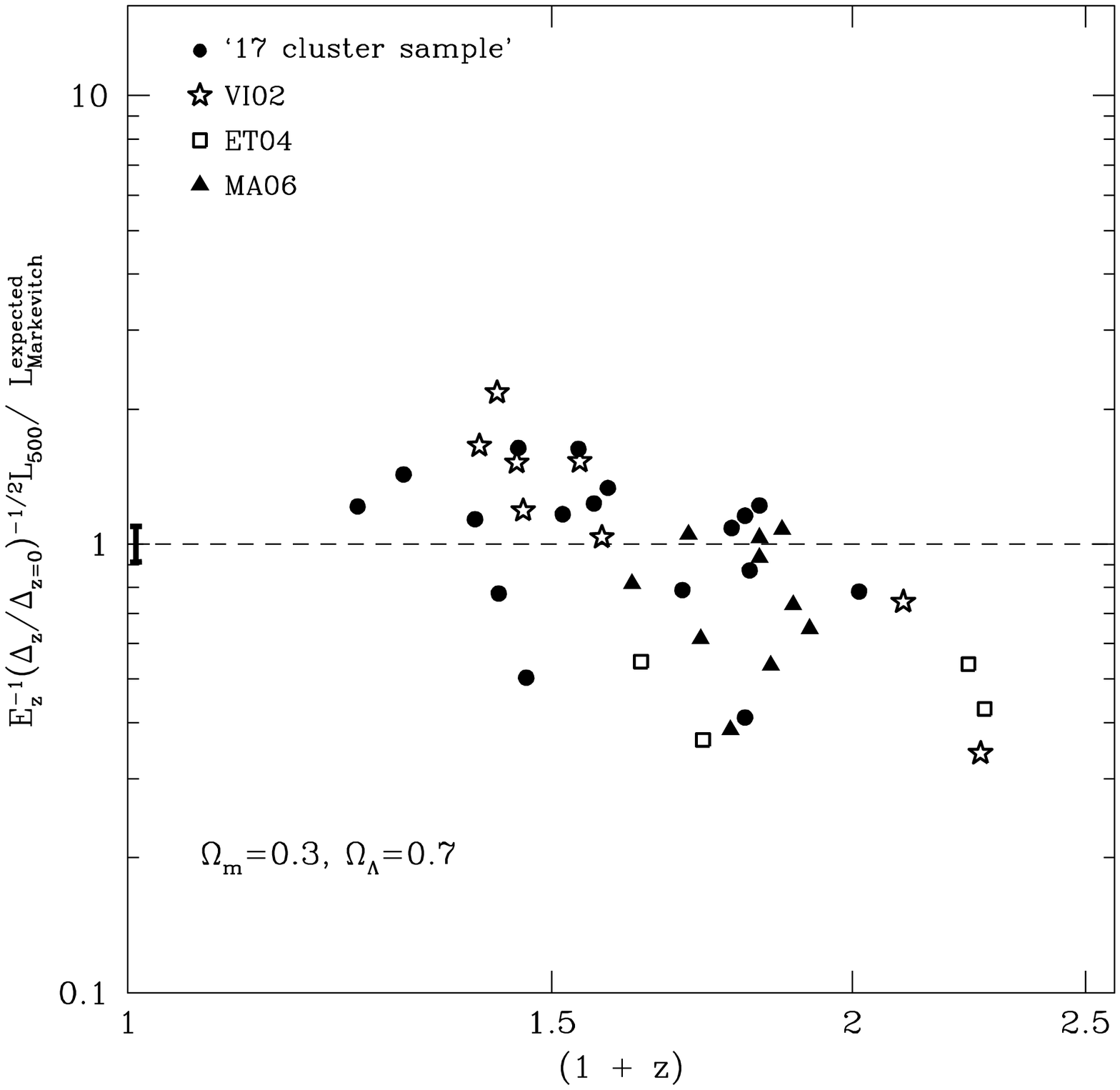}\includegraphics{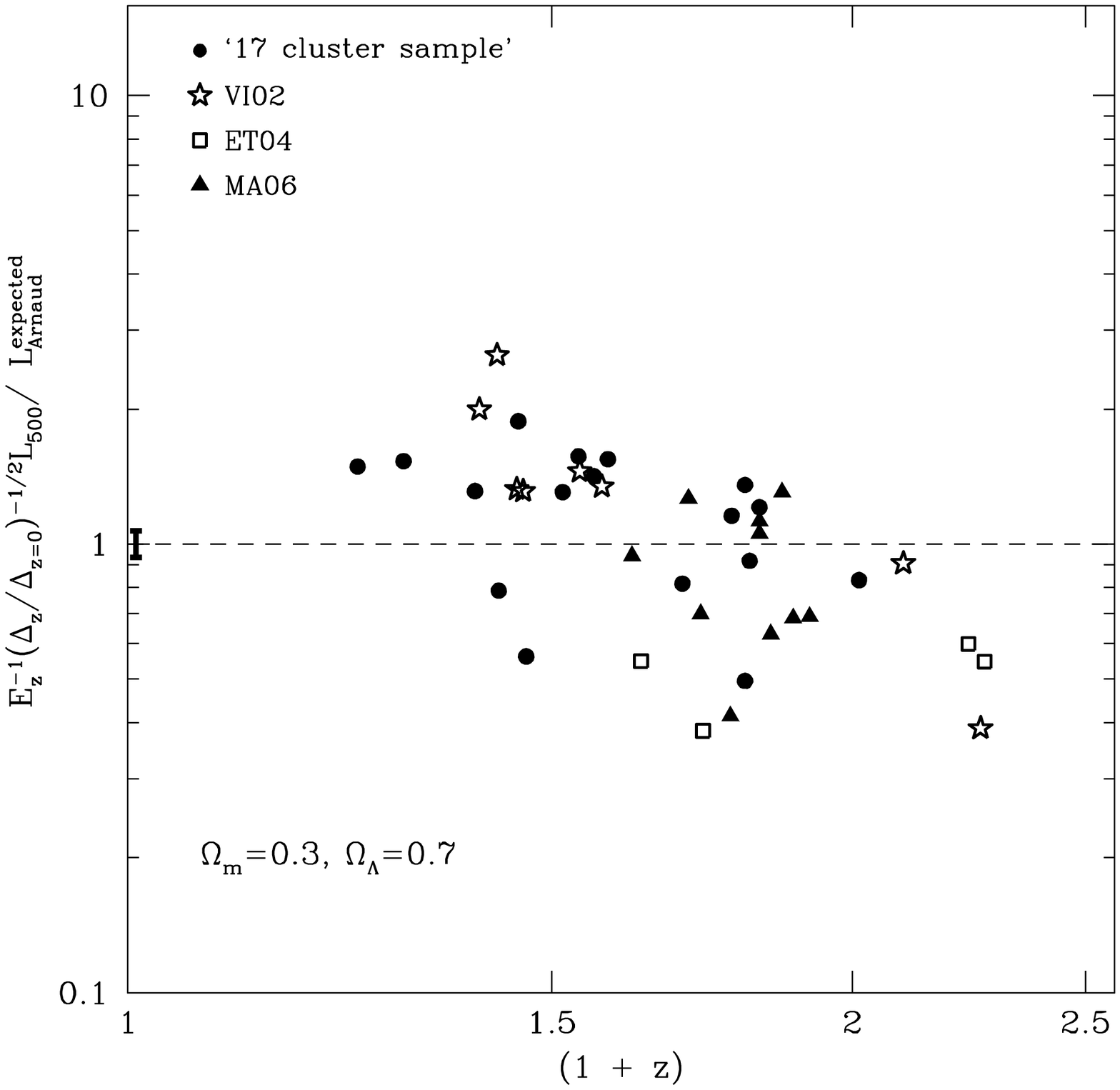}}
\end{center}
\caption{ 
({\it Left panel}) Plot of the ratio between the luminosity scaled by
the cosmological factor $E_z^{-1}
\left(\Delta_z/\Delta_{z=0}\right)^{-1/2}$ and the `expected
luminosity' corresponding to the cluster temperature according to the
local relation of \cite{Ma98} versus $(1+z)$ for the `combined
sample'.  ({\it Right panel}) Same as above but for the local
relation of \cite{Ar99}. In both panels the error bar at z=0 
indicates the uncertainty on the normalization of the local sample 
best-fit law.}
\label{fig4}
\end{figure}

\medskip\noindent
In Table~\ref{tab4} we list the results of the fits obtained by
scaling the luminosities by the cosmological factor $E_z^{-1}
\left(\Delta_z/\Delta_{z=0}\right)^{-1/2}$, as described in step {\it
d)}, and fixing $A_z = 0$.  The best fit values of parameters $\alpha$
and C are found to be very similar to those of the low redshift
clusters.  This result is illustrated in Fig.~\ref{fig3} where the
high redshift cluster data points, corrected for the cosmological
factor, are randomly distributed on the low-z relations
(see also ET04 and MA06).  However the $\chi^2$ value in
Table~\ref{tab4} for the `combined sample' shows that the fit is not
good (probability lower than 1\% to be acceptable).  
This is emphasized in Figure~\ref{fig4} that shows the
ratios between the luminosity scaled by the cosmological factor
$E_z^{-1}\left(\Delta_z/\Delta_{z=0}\right)^{-1/2}$ and the luminosity
corresponding to the cluster temperature according to the local
relations as a function of (1+z) for the `combined sample'. If the
evolution were consistent with the predictions of the self--similar
models, then the data points should be randomly distributed around
unity, independent of the redshift. Figure~\ref{fig4} shows instead an
excess of clusters above unity for $z \le 0.6$ and a large excess
below unity for z $>$ 0.6.

\smallskip\noindent 
When $A_{z}$ is let free to vary we find a strong negative Ev$_1$ evolution 
with the normalization factor C increasing by $\approx$ a factor 2 (see
Table~\ref{tab4}) which brings the data points to shift back,
away from the local relations.  This is the same effect mentioned
earlier. In order to bring the high-z clusters data points in
agreement with the local relation a redshift dependent correction must
be applied to the data. This effect is not present in the redshift
range explored by the combined sample of 39 clusters. 

\bigskip\noindent
We are led to think that the dependence on redshift of the $\rm
L_{bol}-T$ relation in the full range $0 <$ z $< 1.3$ cannot be
described by either the factor
$E_z\left(\Delta_z/\Delta_{z=0}\right)^{1/2}$ or by a power law factor
$(1+z)^A$. These contradictions between data and evolution models
(Ev$_1$ and Ev$_2$) were already noted by \cite{Vo05b} who used a
collection of published data. Voit pointed out that the luminosity
evolution can not be monotonic with redshift. He suggested that the
self--similar evolution may be plausible at low redshift, but cannot
continue to arbitrarily high redshift. This might be consistent with
radiative cooling and feedback from galaxy formation being
increasingly more important at high redshift.
%----------------------------Figure 4-----------------------------
%-----------------------------------------------------------------

\smallskip\noindent
Unless there are systematic observational effects between the low-z
and high-z cluster data (for instance a systematic temperature
difference of $\approx 25$\% or an appropriate combination of
systematic differences in luminosity and temperature) our results
require a strong evolution, similar or stronger than the self--similar, from
z = 0 to z $\le 0.3$ followed by a null evolution at higher redshift.
Systematic differences in both luminosity and temperature
between past and current generation of X-ray detectors are not reported
in the literature. \cite{Vi02} specifically verified that no bias is
present between the temperatures measured by {\em Chandra} and by {\em
ASCA}.

\smallskip\noindent
An additional effect could be the presence of the Malmquist
bias in flux-limited samples \citep[Sect.~5.2 in][]{Ma07a}. However,
this effect is very difficult to quantify with the existing heterogeneous
archival cluster samples.

%-------------------------------subsection----------------------------
\subsection{Comparison with additional Chandra data}

\begin{figure}
\includegraphics[width=10cm,height=10cm,bb=-210 120 364 694]
                {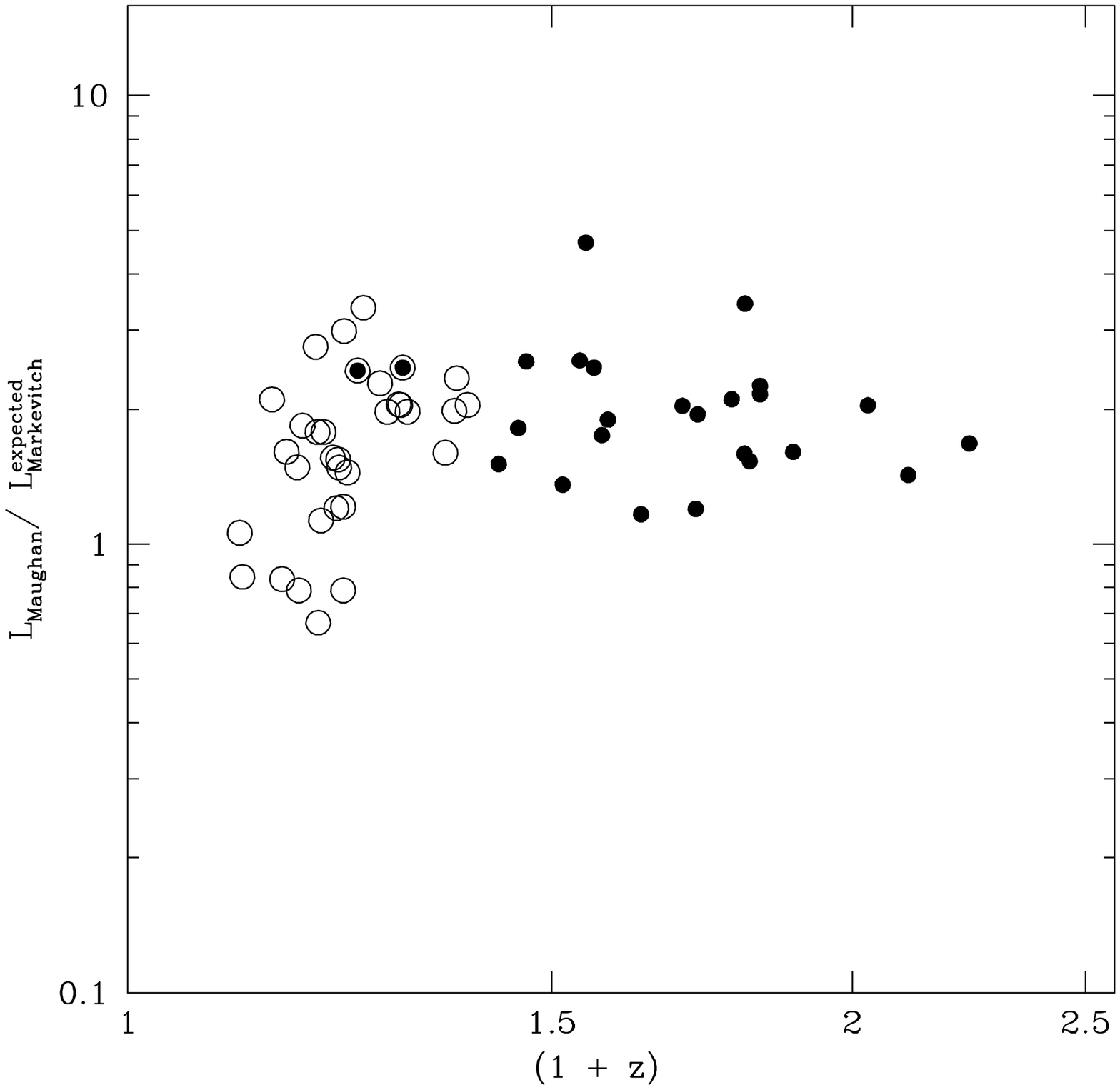}
\vspace{-0.5cm}
\caption{ 
Ratio of the luminosity measured by \cite{Ma07b}  to the `expected
luminosity'  derived using the \cite{Ma07b} measured temperature 
in the local relations by \cite{Ma98} for the 56 clusters (see text
for details). Filled circles indicate the 25 objects  in common with 
our `combined sample'. Open circles are objects with z $<$ 0.4 and 
with a ratio between the luminosities with and without the central
cluster region  less than 1.7 \citep[see columns 4 and 8 of Table 3
in][]{Ma07b}.}
\label{fig5}
\end{figure}

A very recent work by \cite{Ma07b} analyzes a sample of 115 galaxy
clusters at 0.1 $<$ z $<$ 1.3 observed with {\em Chandra}.  It is
interesting for us to use the \cite{Ma07b} sample since they have a
large number of nearby objects (z $<$ 0.4) which are missing in our
sample.  From their 115 clusters we selected 56 clusters as follows.
Twenty-five objects are in common with our `combined sample' clusters
and, according to us, are not contaminated by the central ``cooling
core''. These 25 clusters were used to verify the presence of
systematic difference between ours and Maughan measurements.  Our
estimates of the cluster temperature and luminosity are on average
15\% and 10\% higher with respect to \cite{Ma07b}. We do not discuss
here the possible reasons for these systematic differences, but just
take them into account in the analysis.  The remaining clusters were
selected from the z $<$ 0.4 sample and are those clusters for which
the ratio between the luminosities with and without the central
cluster region is less than 1.7 \citep[see columns 4 and 8 of Table 3
in][]{Ma07b}. In such a way we exclude clusters with strong cooling
cores. Figure~\ref{fig5} is the same as Fig.~\ref{fig2} (left panel)
but for the Maughan's cluster measurements. A comparison between
Fig.~\ref{fig5} and Fig.~\ref{fig2} shows that the two distributions
are very similar\footnote{The systematic differences in luminosity and
temperature mentioned above convert into a higher normalization (about
25\%) in the y-axis of the Maughan distribution with respect to
ours.} and confirms and extends to lower redshift our results. 
In fact the z $>$ 0.25 clusters have on average a flat distribution 
higher than $\rm L_{Maughan}/L_{Markevitch}^{expected} = 1 $ which
is the local cluster value.  At lower redshift (z $<$ 0.25) the region 
pertaining to the local samples begins to being suddenly populated 
(8 clusters). The reason of this change in the distribution is not 
at all clear and not accounted for to our knowledge by any model.

%----------------------------Figure 5-----------------------------

%----------------------------- section ----------------------------
\subsection{Comparison with non--gravitational models} 

\begin{figure}
\begin{center}
\includegraphics[width=13cm,height=12.7cm,bb=50 150 602 690]{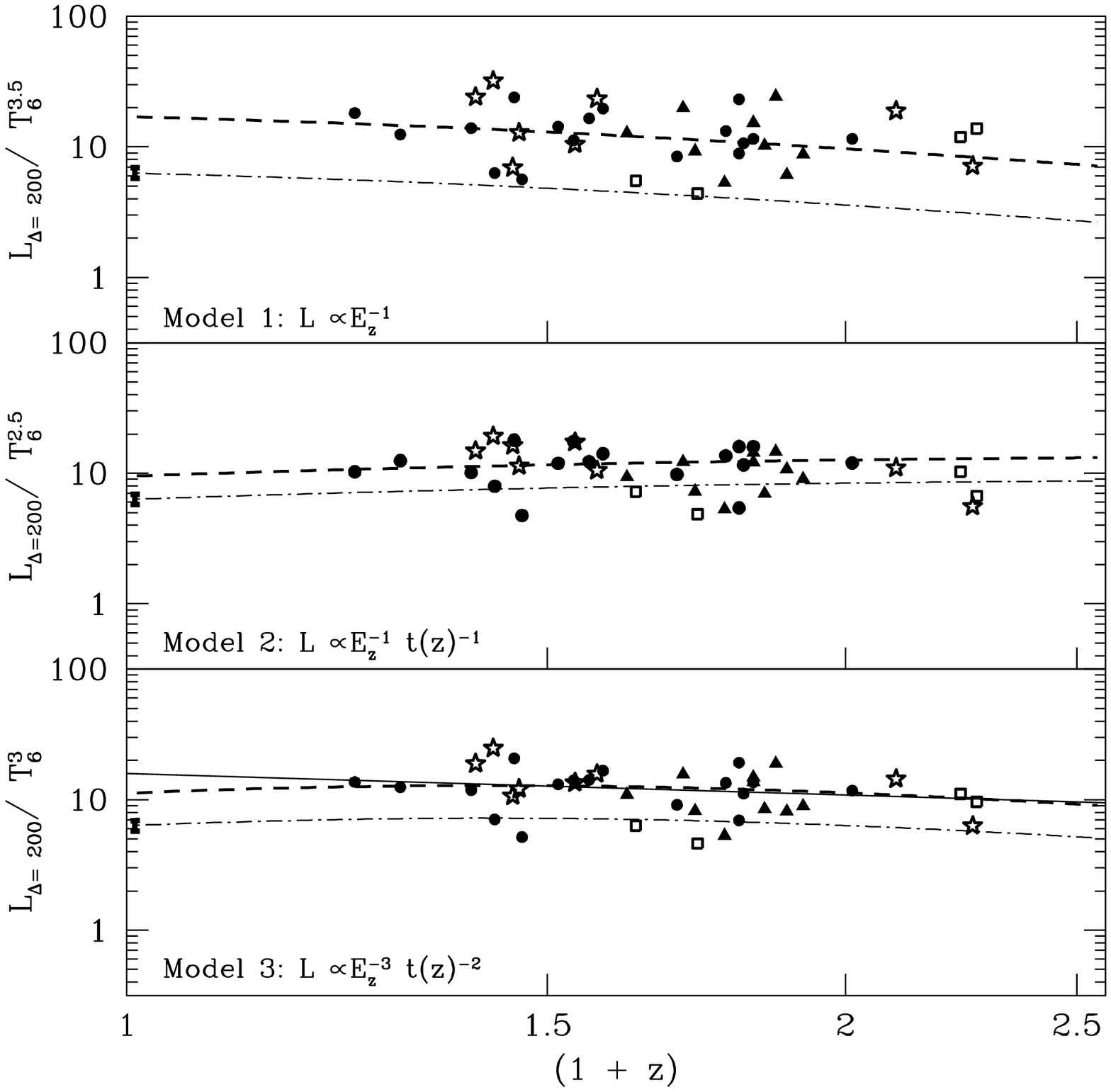}
\end{center}
\vspace{-0.5cm}
\caption{
Ratio between the $\rm L_{\Delta = 200}$ and $\rm T^{\alpha}$, where
$\alpha$ is the temperature dependence of the $\rm L_{bol}-T$ relation
found by \cite{Vo05a}, versus $\rm (1+z)$.  The dashed lines indicate
the redshift dependence of the non-gravitational models, considering only
the `combined sample'.  The dot-dashed lines represents the
non-gravitational models when they are normalized to the local samples.
The solid line in the bottom panel represents the fit obtained
following step {\it b)} described in Sect.~\ref{ltanaly} and using
$\rm L_{\Delta = 200}$ (A $= -0.55\pm0.66$, $\alpha =3.00\pm0.22$,
log C $= 1.20\pm0.13 $, $\chi^2_{min}/d.o.f. = 1.01 $). The error bar at 
z=0 in each panel indicates the normalization of the local samples.}
\label{fig6}
\end{figure}

Figure~\ref{fig6} shows the comparison between the high-z `combined sample' 
and the three non-gravitational models described by  \cite{Vo05a} where 
the effects of preheathing and radiative cooling are included in the simple 
self-similar scenario. To properly compare our data to the Voit models
the luminosities were extrapolated to a radius corresponding to a fixed 
overdensity ($\Delta = 200$) with respect to the critical density as 
required by the models. The three panels in Fig~\ref{fig6} give the ratio 
of the extrapolated luminosity to the $\rm T^{\alpha}$, where
$\alpha$ is the slope of the $\rm L_{bol}-T$  in each model versus (1$+$z).
The dashed lines represent the fit of the models to the data, the dot-dashed 
lines represent the case when the models are normalized to the local 
samples at z$=$0.

\smallskip\noindent
It is worth noting that if only the `combined sample' (0.25 $\le$ z $\le$ 1.3)
is considered the three models (dashed lines in Fig.~\ref{fig6})
are in agreement with the cluster data. The model that better describes
the redshift evolution of the $\rm L_{bol}-T$ relation is the third model 
( $ {\rm L_{bol} \propto T^{3}} (E_z)^{-3} t(z)^{-2}$) shown
in the bottom panel of Fig.~\ref{fig6}. The corresponding minimum $\chi^2$
has a probability of about 50\% to be acceptable ($\chi^2_{min}/d.o.f.=37.3/38$) 
against a probability of about 5\% for the other models  (model 1 
$\chi^2_{min}/d.o.f.=51.8/38$, model 2 $\chi^2_{min}/d.o.f.=57.3/38$).
The third model has also the same temperature slope ($\alpha = 3$) as
the fit of a power law ($\rm L_{\Delta = 200} = C {\rm T_{6}^{\alpha}} (1+z)^A$)
to the data (solid line in the bottom panel of Fig~\ref{fig6}). 

\smallskip\noindent
When we link the high-z `combined sample' to the local samples (see
dot-dashed lines in Fig.~\ref{fig6}) none of these model is
acceptable. Namely none of the non-gravitational models seems to be
able to account for the redshift dependence in the entire redshift
range 0 $\le$ z $\le$ 1.3.

%----------------------------Figure 6-----------------------------
%------------------------------------------------------------------------
\section{Summary and Conclusions}

In this paper we have presented a re-visitation of the observational
relation between X--ray luminosity and temperature for high--z
clusters of galaxies. To this purpose we have constructed the
L$_{\rm bol}$--T relation for a sample of 17 distant clusters (0.25 $<$ z
$<$ 1.01) selected from the {\it Chandra} archive. A detailed analysis
of the X-ray data are given in a companion paper \citep[Paper I by][]{Br07a}. 
The statistics was increased by adding to our sample 22
clusters taken from the literature to form a final sample of 39
objects in the redshift range 0.25 $<$ z $<$ 1.3, observed either by
{\em Chandra} or XMM--{\em Newton}. The analysis of the larger sample
allows a more stringent determination of the high--z L$_{\rm bol}$--T
relation for a comparison with the local one and with the theoretical
models. We emphasize the importance to extrapolate the luminosities to
a fixed fraction of the virial radius when observational results are
compared to theoretical predictions.  The present analysis indicates
that the apparently different results on the L$_{\rm bol}$--T found by
several authors can be in some cases explained by the different
methods used, for instance a different choice of the cluster radius, a
different definition of evolution (Ev$_1$ or Ev$_2$) and/or a
different way to compare the observational results with the
self--similar predictions.

\smallskip\noindent
Our analysis of the L$_{\rm bol}$--T relation of high redshift clusters
has revealed a significant evolution with respect to the local
Universe.  The evolution goes in the direction of high-z clusters
being brighter than the local ones by a factor $\approx 2$ at any
given temperature.  The slope of the relationship is found to be
steeper than expected from the self--similar model predictions and 
steeper, even though still compatible within the errors, than
the local slope \citep{Ma98,Ar99}.

\smallskip\noindent 
For the analysis of the L$_{\rm bol}$--T evolution with redshift,
different approaches were adopted (see Section \ref{ltanaly}):\\ 
-- the data corresponding to the distant clusters in the  redshift 
   range 0.25 $\le$ z $\le$ 1.3 were fitted considering the
   evolution factor (1+z)$^A$;\\ 
-- a second fit is performed with the evolution factor (1+z)$^A$ 
   and fixing the slope and normalization to the best-fit parameters 
   obtained for the well studied low-z cluster relations;\\ 
-- the luminosities are corrected for the cosmological factor predicted by 
   the self--similar model $\rm
   E_z^{-1} \left(\Delta_z/\Delta_{z=0}\right)^{-1/2}$ before the fit is
   performed.

\smallskip\noindent
The following results are found:\\ 
{\it i)} The high-z `combined sample' shows no significant changes
with z in the explored range 0.3 $\le$ z $\le$ 1.3. The fit
of the data with a power law $(1+z)^A$ yields for A a value not
significantly different from zero.\\ 
{\it ii)} The self--similar model is not consistent with the data.\\ 
{\it iii)} When the high-z `combined sample' is linked to the local
samples no acceptable fit is obtained either using a power law $(1+z)^A$ 
or the self--similar scaling law.

\smallskip\noindent
The data from a recent paper by \cite{Ma07b} confirm our results
down to z $\backsimeq 0.25$. In addition, we note in the range 
0.1 $<$ z $<$ 0.25 the appearance of a number of clusters with the 
same properties as the local clusters.

\smallskip\noindent
The non-gravitational models by \cite{Vo05a} are in good agreement with the data
when the redshift range 0.3 $<$ z $<$ 1.3 is considered. When the entire range 
0 $<$ z $<$ 1.3 is taken into account none of these models is
acceptable.

\smallskip\noindent
Summarizing we find that the dependence of the L$_{X}$--T relation on
redshift in the full redshift range explored (0 $\le$ z $\le$ 1.3)
cannot be described by the factor $\rm E_z
\left(\Delta_z/\Delta_{z=0}\right)^{1/2}$ of the self--similar model,
or by any power law of the form $(1+z)^A$. It seems that a strong
evolution is required by the L$_{X}$--T relation from z $=$ 0 to z
$\sim$ 0.3 followed by a much weaker, if any, evolution at
higher redshift. This weaker evolution is compatible with an
increasing importance at high redshift of non-gravitational
effects in the structure formation process \citep{Vo05b}.

\medskip\noindent
According to us the first priority for a correct analysis of the 
evolution of the L$_{\rm bol}$--T relation is to assemble a more 
statistically significant sample of nearby (z $<$ 0.25) clusters. 
A handful of clusters not contaminated by cooling cores and
at z $<$ 0.1 could be enough to shed light on this issue. 
This should be kept in mind for the future generation satellites which 
should have a larger field of view and the same or better resolution 
and sensitivity  than the present X-ray telescopes.

\smallskip\noindent
An additional effect to take into account since it could mimic evolution is
the presence of the Malmquist bias. With the available archival samples it is 
very difficult to keep under control and factor out this bias. It would be
extremely useful to observe with {\em Chandra} or XMM-{\em Newton} 
complete samples of nearby and  distant clusters all coming from the same
survey with well defined selection criteria  in order to be able to control 
and eventually correct for the bias.

%-------------------Acknowledgments----------------------------------
\begin{acknowledgements}
We acknowledge stimulating discussions with Stefano Ettori, Ben
Maughan, Paolo Tozzi and Anna Wolter. We are grateful to an anonymous
referee whose comments contributed to improve this paper. This
research made use of data obtained from the Chandra Data Archive,
which is part of the Chandra X-Ray Observatory Science Center,
operated for the National Aeronautics and Space Administration (NASA)
by the Smithsonian Astrophysical Observatory. Partial financial
support for this work came from the Italian Space Agency ASI (Agenzia
Spaziale Italiana) through grant ASI-INAF I/023/05/0.
\end{acknowledgements}
%-----------------------------------------------------
\clearpage
%
%---------------------------Table 2------------------------------------------
\begin{table}
\caption{Best fit results of $\rm L_{bol}-T$ relation -- Luminosities 
extrapolated to $\rm R_{500}$ (steps {\it a} and {\it b})} 
\begin{center}
\begin{tabular}{lcrrr}
\hline
\\
sample   &  A  & $\alpha $    & log C    & $\chi^2_{min}/d.o.f.$\\ 
\hline
\hline
\\
17 cluster sample  & 0 & 3.14$^{~+~0.34}_{~-~0.29}$ & 
1.09$^{~+~0.04}_{~-~0.04}$  &  12.18/15\\
\\	   
  &--~0.08$^{~+~0.98}_{~-~1.22}$& 3.15$^{~+~0.57}_{~-~0.40}$ &
1.10$^{~+~0.22}_{~-~0.18}$& 12.15/14\\
\hline	 
\\
combined sample  & 0 & 3.00$^{~+~0.19}_{~-~0.17}$ & 
1.06$^{~+~0.03}_{~-~0.03}$  & 39.29/37\\
   &      &               &                                &\\       
  & --~0.48~$^{~+~0.58}_{~-~0.73}$& 3.03$^{~+~0.25}_{~-~0.20}$ 
		 & 1.15$^{~+~0.15}_{~-~0.10}$  & 37.18/36                
\\
\hline
\label{tab2}
\end{tabular}
\vspace{-0.5cm}
\end{center}
\end{table} 
%-------------------------------------------------------------------
%-------------------------Table 3----------------------------------------
\begin{table}
\caption{Best fit results of $\rm L_{bol}-T$ relation -- Luminosities 
extrapolated to $\rm R_{500}$ (step {\it c})} 
\begin{center}
\begin{tabular}{ccrcc}
\hline
\\
sample                &  A  & $\chi^2_{min}/d.o.f.$ & Reference sample\\ 
\hline
\hline
\\
17 cluster sample   &$1.40^{~+~0.13}_{~-~0.13}$ &  22.39/16 & \cite{Ma98}  &\\ 
\\                      
	      &$1.58^{~+~0.13}_{~-~0.15}$ &  23.26/16 & \cite{Ar99}  &\\ 
\hline
\\
combined sample   &$1.20^{~+~0.08}_{~-~0.08}$ &  73.57/38 & \cite{Ma98}  &\\ 
\\                     
             &$1.35^{~+~0.08}_{~-~0.10}$ & 68.35/38 & \cite{Ar99}  &\\ 
\hline 
\label{tab3}
\end{tabular}
\vspace{-0.5cm}
\end{center}     
The slope ($\alpha$) and the normalization (log C) are fixed to the best-fit 
values obtained for the low-z cluster sample indicated in the last column.
\end{table}
%----------------------------------------------------------------------------
%---------------------------Table 4--------------------------------------
\begin{table}
\caption{Best fit results of $E_z^{-1}
\left(\Delta_z/\Delta_{z=0}\right)^{-1/2} {\rm L_{bol}-T}$ relation 
(step {\it d})} 
\begin{center}
\begin{tabular}{lcccr}
\hline
sample    & $A_z$  & $\alpha $   & log C       & $\chi^2_{min}/d.o.f.$\\ 
\hline
\hline
\\
17 cluster sample & 0&2.93$^{~+~0.33}_{~-~0.28}$ &
0.89$^{~+~0.04}_{~-~0.04}$  &17.99/15\\
\\	   
            &--~1.28$^{~+~0.98}_{~-~1.22}$&3.17$^{~+~0.55}_{~-~0.43}$
            &1.13$^{~+~0.23}_{~-~0.18}$& 12.18/14\\
\hline
\\
combined sample   &  0 &3.00$^{~+~0.19}_{~-~0.18}$ &
0.87$^{~+~0.03}_{~-~0.03}$& 61.09/37 \\
\\	   
               &--~1.80$^{~+~0.70}_{~-~0.65}$&3.05$^{~+~0.23}_{~-~0.23}$
	 & 1.20$^{~+~0.13}_{~-~0.13}$ & 37.37/36\\
\hline	 
\label{tab4}
\end{tabular}
\vspace{-0.5cm}
\end{center}
The luminosities extrapolated to $\rm R_{500}$ correspond
to a redshift-dependent contrast of 500, and were scaled by the cosmological 
factor $E_z^{-1} \left(\Delta_z/\Delta_{z=0}\right)^{-1/2}$ as predicted 
by the self--similar model.
\end{table} 
%-----------------------------------------------------
%---------------------------References----------------------------------


\clearpage
\begin{thebibliography}{}
\bibitem[Allen(2003)]{Al03} Allen, S.~W.\ 2003, \apss, 285, 
  247 
\bibitem[Arnaud \& Evrard(1999)]{Ar99} Arnaud, M., \& 
  Evrard, A.~E.\ 1999, \mnras, 305, 631 
\bibitem[Balogh et al.(1999)]{Bal99} Balogh, M.~L., Babul, 
  A., \& Patton, D.~R.\ 1999, \mnras, 307, 463 
\bibitem[Borgani et al.(2002)]{Bor02} Borgani, S., Governato, 
  F., Wadsley, J., Menci, N., Tozzi, P., Quinn, T., Stadel, J., \& Lake, G.\ 
  2002, \mnras, 336, 409 
\bibitem[Borgani et al.(2004)]{Bo04} Borgani, S., et al.\ 
  2004, \mnras, 348, 1078
\bibitem[Borgani (2006)]{Bo06}  Borgani, S., 2006, astro-ph/0605575
\bibitem[Bower et al.(2001)]{Bow01} Bower, R.~G., Benson, 
   A.~J., Lacey, C.~G., Baugh, C.~M., Cole, S., \& Frenk, C.~S.\ 2001, \mnras, 
   325, 497
\bibitem[Branchesi et al.(2007a)]{Br07a} Branchesi, M., Gioia, 
  I.~M., Fanti, C., \& Fanti, R.\ 2007a, ArXiv e-prints, 706, 
  arXiv:0706.3272 (Paper I) 
\bibitem[Branchesi et al.(2007b)]{Br07b} Branchesi, M., Gioia, 
  I.~M., Fanti, C., Fanti, R., \& Cappelluti, N.\ 2007b, \aap, 462, 449 
\bibitem[Brighenti \& Mathews(2001)]{BM01} Brighenti, F., \& 
  Mathews, W.~G.\ 2001, \apj, 553, 103
\bibitem[Bryan \& Norman(1998)]{Bry98} Bryan, G.~L., \& 
  Norman, M.~L.\ 1998, \apj, 495, 80 
\bibitem[Bryan(2000)]{Bry00} Bryan, G.~L.\ 2000, \apjl, 544, L1 
\bibitem[Cavaliere \& Fusco-Femiano(1976)]{Ca76} Cavaliere, 
  A., \& Fusco-Femiano, R.\ 1976, \aap, 49, 137
\bibitem[Ettori et al.(2004)]{Et04} Ettori, S., Tozzi, P., 
  Borgani, S., \& Rosati, P.\ 2004, \aap, 417, 13 (ET04)
\bibitem[Evrard \& Henry(1991)]{Ev91} Evrard, A.~E., \& 
  Henry, J.~P.\ 1991, \apj, 383, 95
\bibitem[Evrard et al.(1996)]{Ev96} Evrard, A.~E., Metzler, 
  C.~A., \& Navarro, J.~F.\ 1996, \apj, 469, 494 
\bibitem[Evrard et al.(2002)]{Ev02} Evrard, A.~E., et al.\ 
  2002, \apj, 573, 7 
\bibitem[Fabian et al.(1994)]{Fa94} Fabian, A.~C., Crawford, 
  C.~S., Edge, A.~C., \& Mushotzky, R.~F.\ 1994, \mnras, 267, 779 
\bibitem[Kaiser(1986)]{Ka86} Kaiser, N.\ 1986, \mnras, 222, 
  323 
\bibitem[Kaiser(1991)]{Ka91} Kaiser, N.\ 1991, \apj, 383, 
  104 
\bibitem[Kotov \& Vikhlinin(2005)]{Ko05} Kotov, O., \& 
  Vikhlinin, A.\ 2005, \apj, 633, 781 
\bibitem[Markevitch(1998)]{Ma98} Markevitch, M.\ 1998, \apj, 
  504, 27 
\bibitem[Maughan et al.(2006)]{Ma06} Maughan, B.~J., Jones, 
  L.~R., Ebeling, H., \& Scharf, C.\ 2006, \mnras, 365, 509 (MA06)
\bibitem[Maughan(2007a)]{Ma07a} Maughan, B.~J.\ 2007a, ArXiv 
Astrophysics e-prints, arXiv:astro-ph/0703504 
\bibitem[Maughan et al.(2007b)]{Ma07b}
 Maughan, B.~J., Jones, C., Forman, W., \& Van Speybroeck, L.\ 2007b, 
 ArXiv Astrophysics e-prints, arXiv:astro-ph/0703156 
\bibitem[Muanwong et al.(2006)]{Mu06} Muanwong, O., Kay, 
  S.~T., \& Thomas, P.~A.\ 2006, \apj, 649, 640 
\bibitem[Ponman et al.(1999)]{Po99} Ponman, T.~J., Cannon, 
  D.~B., \& Navarro, J.~F.\ 1999, \nat, 397, 135 
\bibitem[Reiprich \& B{\"o}hringer(2002)]{Rei02} Reiprich, 
  T.~H., B{\"o}hringer, H.\ 2002, \apj, 567, 716 
\bibitem[Spitzer (1978)]{Sp78} Spitzer, L., Jr. 1978, Physical 
  Processes in the Interstellar Medium (New York: Wiley)
\bibitem[Tozzi \& Norman(2001)]{Tn01} Tozzi, P., \& Norman, 
  C.\ 2001, \apj, 546, 63 
\bibitem[Tozzi(2006)]{Toz06} Tozzi, P.\ 2006, ArXiv 
  Astrophysics e-prints, arXiv:astro-ph/0602072 
\bibitem[Tornatore et al.(2003)]{Tor03} Tornatore, L., 
  Borgani, S., Springel, V., Matteucci, F., Menci, N., \& Murante, G.\ 2003, 
  \mnras, 342, 1025 
\bibitem[Vikhlinin et al.(2002)]{Vi02} Vikhlinin, A., 
  VanSpeybroeck, L., Markevitch, M., Forman, W.~R., \& Grego, L.\ 2002, 
  \apjl, 578, L107 (VI02)
\bibitem[Voit \& Bryan(2001)]{Vo01} Voit, G.~M., \& Bryan, 
  G.~L.\ 2001, \nat, 414, 425
\bibitem[Voit et al.(2002)]{Vo02} Voit, G.~M., Bryan, G.~L., 
  Balogh, M.~L., \& Bower, R.~G.\ 2002, \apj, 576, 601
\bibitem[Voit(2005a)]{Vo05a} Voit, G.~M., 2005a, Reviews of Modern Physics 
  77, 207
\bibitem[Voit(2005b)]{Vo05b} Voit, G.~M.\ 2005b, Advances in 
  Space Research, 36, 701 
\bibitem[Wu \& Xue(2002)]{Wu02} Wu, X.-P., \& Xue, Y.-J.\ 
  2002, \apj, 569, 112 
\end{thebibliography}
\end{document}